%% file: ms.tex
\documentclass[12pt,preprint]{aastex}
\usepackage{epsf}
\input{defs.tex}

\begin{document}

\title{Numerical Simulations of Gaseous Disks Generated from Collisional 
Cascades at the Roche Limits of White Dwarf Stars}
\vskip 7ex
\author{Scott J. Kenyon}
\affil{Smithsonian Astrophysical Observatory,
60 Garden Street, Cambridge, MA 02138} 
\email{e-mail: skenyon@cfa.harvard.edu}

\author{Benjamin C. Bromley}
\affil{Department of Physics \& Astronomy, University of Utah, 
201 JFB, Salt Lake City, UT 84112} 
\email{e-mail: bromley@physics.utah.edu}
%
%

\begin{abstract}

We consider the long-term evolution of gaseous disks fed by the vaporization
of small particles produced in a collisional cascade inside the Roche limit 
of a 0.6~\msun\ white dwarf. Adding solids with radius \r0\ at a constant rate 
\mdotz\ into a narrow annulus leads to two distinct types of evolution. When 
\mdotz\ $\gtrsim \dot{M}_{0, crit}$ $\approx$ $3 \times 10^4 ~ (r_0 / {\rm 1~km})^{3.92}$~\gs,
the cascade generates a fairly steady accretion disk where the mass transfer rate
of gas onto the white dwarf is roughly \mdotz\ and the mass in gas is
$M_g \approx 2.3 \times 10^{22} ~ (\mdotz / 10^{10}~\gs) ~ ({\rm 1500~K} / T_0) ~ (10^{-3} / \alpha)$~g, 
where $T_0$ is the temperature of the gas near the Roche limit and 
$\alpha$ is the dimensionless viscosity parameter.
If \mdotz\ $\lesssim \dot{M}_{0, crit}$, the system alternates between high states with 
large mass transfer rates and low states with negligible accretion.  Although either 
mode of evolution adds significant amounts of metals to the white dwarf photosphere,
none of our calculations yield a vertically thin ensemble of solids inside the Roche
limit. X-ray observations can place limits on the mass transfer rate and test this 
model for metallic line white dwarfs.

\end{abstract}


\section{INTRODUCTION}
\label{sec: intro}

Over the past four decades, observations have shown that many white dwarfs have 
metallic absorption lines from O, Mg, Al, Si, Ca, Fe, and a variety of other elements 
with atomic number $Z \ge$ 6 \citep[e.g.,][and references therein]{shipman1977,
cottrell1980,shipman1983,lacombe1983,liebert1987,kenyon1988,sion1990,zuckerman1998,
jura2014,koester2014,kepler2016,xu2017}. 
Some of these stars have near-IR excess emission from warm dust orbiting near the 
Roche limit 
\citep[e.g.,][]{kilic2005,reach2005,hansen2006,tremblay2007,vonhippel2007,farihi2009,
girven2011, debes2011,chu2011,barber2012,hoard2013,barber2014,bergfors2014,
rocchetto2015,farihi2016,bonsor2017}. 
A few have metallic emission features from ionized or neutral gas, also orbiting 
within the Roche limit \citep[e.g.,][]{gansicke2006,gansicke2007,gansicke2008,
melis2010b,farihi2012,melis2012a,debes2012b,wilson2014}.

Currently popular models for these white dwarfs propose that the photospheric absorption 
lines result from accretion of solid material originally orbiting at large distances 
from the host star \citep[e.g.,][]{alcock1980b,lacombe1983,alcock1986,jura2003,
koester2006,jura2007b,jura2007a,wyatt2014,veras2016}. 
Perturbations of the orbits lead to a succession of solids that
fall within the Roche limit of the white dwarf and eventually form an
optically thick, vertically thin disk surrounding the white dwarf. Vaporization
of disk particles produces a gaseous disk, which moderates direct accretion of 
material onto the white dwarf photosphere \citep[e.g.,][]{debes2002,jura2003,
jura2008,rafikov2011a,debes2012a,metzger2012,veras2013,brown2017}.  

In \citet[][hereafter KB2017]{kb2017b}, we considered the long-term evolution 
of solid material placed on mildly eccentric ($e$ = 0.01) orbits within a 
narrow annulus near the Roche limit of a 0.6~\msun\ white dwarf. Destructive 
collisions generate a collisional cascade which converts 1--100~km asteroids 
into dust grains with radii $r \lesssim$ 1~\mum. When solids are replenished 
at a rate \mdotz, the system often finds an equilibrium which depends on 
\mdotz\ and \r0\ the radius of the largest solid added to the annulus.  
Equilibria with constant mass require \r0\ $\lesssim$ 10--30~km and 
$\mdotz \gtrsim 10^7 - 10^8$~\gs. Otherwise, the solid mass in the annulus
oscillates between high states with large collision rates and low states with 
negligible collision rates.

Throughout all of our simulations, the vertical scale height $H$ of the solids 
remains large, $H \approx 0.01 a$, where $a$ is the distance of the annulus 
from the central star. In principle, collisional damping is sufficient to reduce 
$H$ significantly on 5--10 collision times. In practice, however, the collisional
cascade processes solids on shorter time scales and prevents damping. Thus, 
collisional processes are incapable of assembling a thin disk of solids inside
the Roche limit of a white dwarf. 

Although simple order-of-magnitude estimates suggest that interactions between the 
solids and the gas are also incapable of reducing $H$ (KB2017), it is necessary
to consider whether a more detailed treatment of the gas can produce conditions
more amenable to the formation of a thin disk of solids. Here, we expand on KB2017 
and derive the time evolution of a gaseous disk formed by the vaporization of small
solids produced in the collisional cascade. Once we infer the radial distribution
of the gas surface density ($\Sigma_g$), we use an adopted temperature distribution 
to calculate the impact of the gas on solid particles.

In addition to placing better constraints on the ability of solids to collapse 
into a thin disk, we derive the time evolution of the accretion rate of gas onto
the central white dwarf. These results allow us to begin to compare theoretical
estimates of accretion rates with observations.

After briefly summarizing the algorithms used in our simulations (\S2), we describe 
results for a suite of calculations with different \r0\ and \mdotz\ (\S3).  
We then compare our results with previous investigations, discuss the likely impact 
of the gas on solid particles, and make some initial comparisons with observations (\S4).  
We conclude with a brief summary (\S5).

\section{THEORETICAL BACKGROUND}
\label{sec: back}

To follow the evolution of a gaseous disk generated from vaporized solids,
we rely on \orch, a parallel \verb!C++/MPI! hybrid coagulation + 
\nbody\ code that tracks the accretion, fragmentation, and orbital evolution 
of solid particles ranging in size from a few microns to thousands of km 
\citep{kb2001,kb2004a,kb2008,bk2011a,bk2011b,kb2016a}. 
The ensemble of codes within \orch\ includes a multi-annulus coagulation 
code for small particles, an \nbody\ code for large particles, and 
separate radial diffusion codes for solids and gas.  Several algorithms 
link the codes together, enabling each component to react to the evolution 
of other components.

As in KB2017, we assume solid particles lie within a single annulus of width 
$\Delta a$ at a distance $a_0$ from a central star with mass \mstar\ = 
0.6~\msun\ and radius \rstar\ = 0.013~\rsun\ ($a_0$ = 1.16~\rsun; $\Delta a = 0.2 a_0$).  
Particles on circular orbits have velocities $v_K = (G \mstar\ / a_0)^{1/2}$ 
$\approx$ 300~\kms.  
Within the annulus, there are $M$ mass batches with characteristic mass $m_i$ 
and logarithmic spacing $\delta = m_{i+1} / m_i$; adopting $\delta$ = 1.4 
provides a reasonably accurate solution for the cascade \citep[e.g,][and 
references therein]{kb2015a,kb2015b,kb2016a}.  Batches contain $N_i$ particles 
with total mass $M_i$, average mass $\bar{m}_i = M_i / N_i$, horizontal velocity 
$h_i$ ($e_i$ = $\sqrt{1.6} h_i / v_K$), and vertical velocity $v_i$ 
(sin~$\imath$ = $\sqrt{2} v_i / v_K$).  The number of particles, total mass, 
and orbital velocity of each batch evolve through physical collisions and 
gravitational interactions with all other mass batches in the ring. 

At the start of each calculation, the single annulus is empty of solids. 
During a time step of length $\Delta t$, we add a mass in solids, 
$\delta M = \mdotz\ \Delta t$. Every solid particle added to the annulus
has radius \r0, mass \m0, eccentricity $e_0$, and inclination $\imath_0$.
The solids have a mass density $\rho_s$ = 3~\gcmc.
Along with the input rate \mdotz, the initial properties of the solids are 
held fixed in each calculation.  The number of particles added to the 
grid is $\Delta N = \mdotz\ ~ \Delta t / \m0$. Our algorithm uses a random 
number generator to round $\Delta N$ up or down to the nearest integer.  
For systems with large \r0, this procedure introduces some shot noise into 
the input rate. 

When the solid mass reaches a critical level, the particles begin to collide. 
As summarized in KB2017, the coagulation code within \orch\ derives the rates 
and outcomes of physical collisions and the velocity evolution from gravitational 
interactions among all particles in the grid. By setting $e_0$ = 0.01 and 
$\imath_0 = e_0/2$, we ensure that all collisions are destructive, with approximate 
collision velocities $v_c \approx e_0 v_K$ $\approx$ 3~\kms.  The complete ensemble 
of destructive collisions generates a collisional cascade, where solids with initial 
radii of 1--100~km are gradually ground into 1~\mum\ dust grains.

In KB2017, we assumed that particles with radii $r \lesssim$ 1~\mum\ were
vaporized and `lost' to the system. Here, we consider how this material 
evolves when it feeds a gaseous reservoir. Vaporized solids are placed into 
the reservoir at a rate \mdotv, which is derived from the coagulation code 
every time step.  This material is spread evenly over the width of the annulus, 
which extends from an inner radius \ain\ = $a_0 - \Delta a/2$ to an outer radius 
\aout\ = $a_0 + \Delta a/2$. Between \ain\ and \aout, the reservoir grows in 
surface density at a rate $\dot{\Sigma}_g = \mdotv / 2 \pi a_0 \Delta a$.

As the gaseous reservoir grows, we numerically solve the radial diffusion equation
\citep{lbp1974,pri1981}
\begin{equation}
\frac{\partial \Sigma_g}{\partial t} = 3 a^{-1}~\frac{\partial}{\partial a} ~ \left ( a^{1/2}~\frac{\partial}{\partial a} ~ \{ \nu \Sigma_g a^{1/2} \} \right ) + \left ( \frac{\partial \Sigma_g}{\partial t} \right )_v ~
\label{eq: disk-evol}
\end{equation}
for the evolution of the surface density \citep[see also][]{bk2011a}. Here, $a$ 
is the radial distance from the central star, $\nu$ is the viscosity, and $t$ 
is the time.  The first term is the change in $\Sigma_g$ from viscous evolution; the 
second term is the change in $\Sigma_g$ from vaporization of small solid particles. 

To set the viscosity in each annulus, we adopt a standard prescription
\begin{equation}
\nu = \alpha c_s H_g ~ ,
\label{eq: disk-nu}
\end{equation}
where $\alpha$ is the dimensionless viscosity parameter, $c_s$ is the sound speed,
and $H_g$ is the vertical scale height of the gas. Following \citet{metzger2012}, 
we set $\alpha = 10^{-3}$.  For disk material with angular velocity 
$\Omega = (G \mstar / a_0^3)^{1/2}$, $H_g = c_s \Omega^{-1}$. The sound speed is 
\begin{equation}
c_s = \left ( { \gamma k_B T_g \over \mu m_H } \right )^{1/2} ~ ,
\label{eq: disk-cs}
\end{equation}
where $\gamma = 5/3$ is the ratio of specific heats, $k_B$ is the Boltzmann constant,
$T_g$ is the gas temperature, $\mu$ = 28 is the mean molecular weight, and $m_H$ is 
the mass of a hydrogen atom.

In this application, the energy generated from viscous mass transport is negligible. 
To avoid solving for the physical conditions in the gas \citep[e.g.,][]{melis2010b}, 
we adopt a simple expression for the gas temperature
\begin{equation}
T_g = T_0 \left ( { a  \over \rsun\ } \right )^{-n} ~ ,
\label{eq: disk-temp}
\end{equation}
with $T_0 \approx$ 1500~K and $n$ = 1/2. This expression is similar to the more 
detailed results of \citet{melis2010b}, where $T_0 \approx$ 1500--3000~K and
$n$ = 0.25--0.75 for white dwarfs with effective temperatures $T_{eff}$ = 
5000-15000~K.  To quantify the impact of the adopted $T_0$, we also consider 
evolution of the gas for $T_0$ = 3000~K.

To solve eq.~\ref{eq: disk-evol} with input $\alpha$ and $T_g$, we specify an inner 
radius $a_1$ = 1.5~\rstar\ and an outer radius $a_2$ = 1~AU.  The large outer radius
allows vaporized solids to expand well past the Roche limit.  Setting $x = 2 a^{1/2}$, 
we divide the disk into 1001 annuli equally spaced in $x$ \citep[see also][]{bath1981,
bath1982}. As a standard boundary condition, $\Sigma(a_1) = \Sigma(a_2)$ = 0. Within 
every coagulation time step, our explicit algorithm for the radial surface density 
executes a set of $n$ internal time steps to satisfy the Courant condition and to 
enable mass conservation to machine accuracy.  As a check, we also derive an implicit 
solution for $\Sigma_g(t)$ \citep{press1992}. Over the full range in $a$, the two 
solutions yield the same $\Sigma_g$ to better than 0.1\% over 1--10~Myr of evolution.

At the start of each calculation, the vertical scale height of solids is much larger
than the vertical scale height of the gas. For the solids, $H \approx \imath a$
$\approx$ 7500~km at $a \approx$ 1.15~\rsun. The gas has a vertical scale height
$H_g \approx c_s \Omega^{-1}$ $ \approx $ 1500~km. One of our goals is to learn whether 
collisional processes lead to situations with $H \lesssim H_g$.

\section{RESULTS}

To explore the evolution of a gaseous disk generated by the vaporization of small solids, 
we first consider collisional cascade simulations with \r0\ = 1~km and \mdotz\ = 
$10^7 - 10^{13}$~\gs. The range for \mdotz\ includes accretion rates inferred for 
metallic line white dwarfs \citep{wyatt2014,farihi2016}.  
Calculations with \r0\ = 1~km minimize shot noise, which grows with increasing 
\r0\ (KB2017). 

\begin{figure} 
\includegraphics[width=6.5in]{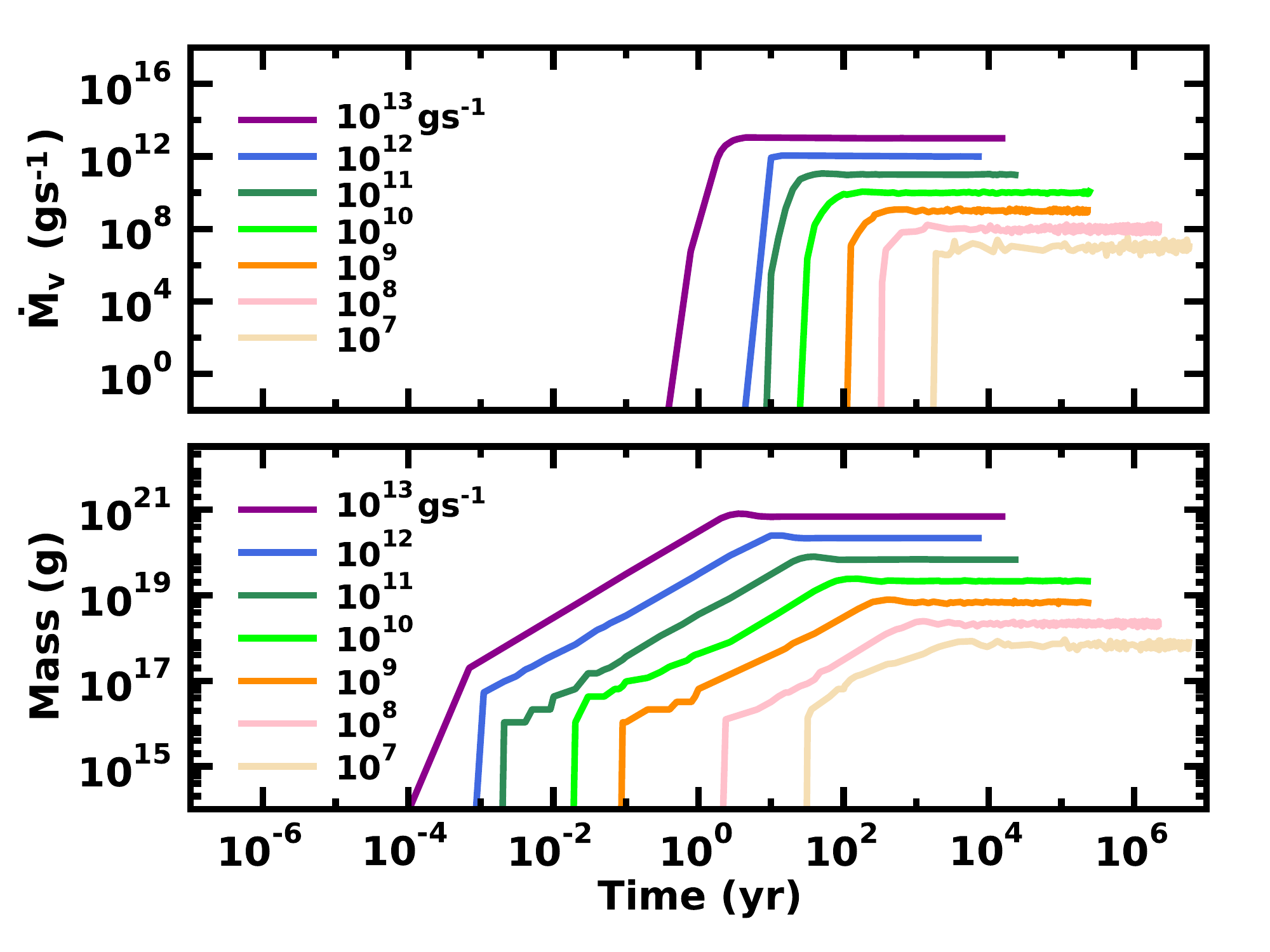}
\vskip 3ex
\caption{
Time evolution of the mass in solids $M_d$ (lower panel) and the mass 
vaporization rate \mdotv\ (upper panel) for calculations with \r0\ = 1~km 
and the mass input rates (\mdotz) indicated in the legend. In each calculation, 
the mass grows roughly linearly in time from zero to a constant level. During 
this increase, the collisional cascade begins; the vaporization rate (equivalently 
the production rate of particles with $r \lesssim$ 1~\mum) grows abruptly and 
then reaches a roughly constant level. A balance between the input rate and the 
vaporization rate maintains a constant mass in the annulus.
\label{fig: mmdot1}
}
\end{figure}

Fig.~\ref{fig: mmdot1} (lower panel) illustrates the evolution of the total mass in solids, 
$M_d$, for various \mdotz\ listed in the caption. At the start of each calculation, the
annulus contains no mass. As mass is added, the collision rate is negligible. Thus,
$M_d$ increases linearly with time. Once the mass reaches a critical limit, destructive
collisions among the solids begin to produce smaller objects. In turn, collisions
between the larger and smaller objects generate even more debris. This process fuels
a collisional cascade which grinds large objects into small dust grains. After many 
collision times, the rate the cascade processes mass equals the input rate \mdotz. 
The mass in solids then achieves a roughly stable value which is maintained for the 
rest of the calculation. In systems with large (small) \mdotz, this mass varies 
slightly (noticeably) with time.

Fig.~\ref{fig: mmdot1} (upper panel) follows the evolution of the vaporization rate \mdotv. 
At the start of the calculation, collisions among large objects are rare; \mdotv\ is 
close to zero. Once the collisional cascade begins, \mdotv\ rises abruptly and then
finds a plateau level where \mdotv\ = \mdotz. For systems with smaller \mdotz, there 
are modest variations in \mdotv\ about the input rate \mdotz.

\begin{figure} 
\includegraphics[width=6.5in]{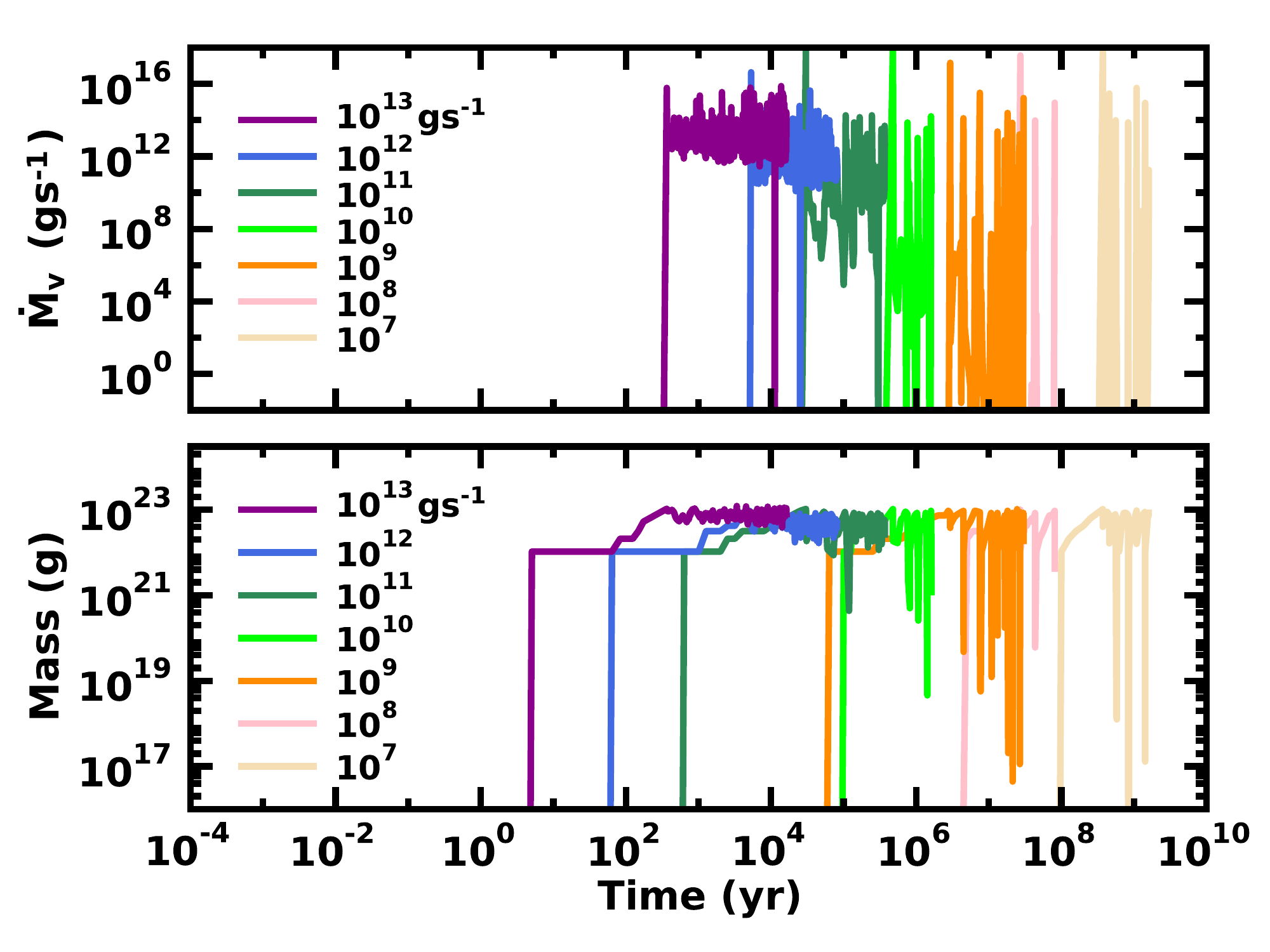}
\vskip 3ex
\caption{
As in Fig.~\ref{fig: mmdot1} for calculations with \r0\ = 100~km. In these
calculations, the mass and vaporization rate either maintain a roughly 
constant level (for large \mdotz) or oscillate between low and high states
(for small \mdotz).
\label{fig: mmdot2}
}
\end{figure}

Fig.~\ref{fig: mmdot2} plots results for calculations with \r0\ = 100~km. When \r0\ is
large, the shot noise in the input rate is also large. The mass of solids in the annulus 
then grows episodically with time (Fig.~\ref{fig: mmdot2}, lower panel). Once the annulus 
contains several large objects, the collisional cascade begins. When \mdotz\ is large,
the cascade processes mass at roughly the same rate as the input rate \mdotz. The mass
in the annulus then exhibits small oscillations about an equilibrium mass which is 
somewhat larger than the equilibrium mass for smaller \r0\ (KB2017). 

For any \mdotz, the number of large objects required to initiate the cascade is 
$N_{min} \gtrsim$ 2.  Systems with smaller \mdotz\ then take longer to start a cascade.
Once the cascade begins, collisions convert large objects into small objects at a rate
that depends only on the mass in the annulus. This rate is independent of \mdotz. For 
systems with smaller \mdotz, collisions process mass at rates much larger than the 
input rate.  The system then oscillates between high states where the collisional 
cascade processes mass rapidly and low states where the system slowly gains enough mass 
to begin a new cascade.

The oscillations in $M_d$ produce similar variations in \mdotv\ (Fig.~\ref{fig: mmdot2}).
Throughout the evolution, \mdotv\ is dominated by shot noise from occasional collisions
among the largest objects. When \mdotz\ is large, there are always enough large objects to
produce a continuous cascade; \mdotv\ then varies slowly about an equilibrium rate comparable
with \mdotz. When \mdotz\ is small, the cascade is intermittent. During the high state, 
\mdotv\ achieves the equilibrium level of systems with large \mdotz. As the system falls into
the low state, \mdotv\ drops to zero.

Systems with different starting values for \r0\ and \mdotz\ behave similarly (KB2017). When
\r0\ is small and \mdotz\ is large, the evolution of the system is very smooth. The cascade
then always adjusts to balance the collision rate with \mdotz; \mdotv\ = \mdotz.
When \r0\ is large and \mdotz\ is small, the evolution is oscillatory. In these systems, 
the cascade cannot find an equilibrium where the rate mass flows down the mass distribution
(from large objects to small objects) and the vaporization rate \mdotv\ equal the mass input 
rate \mdotz. Instead, the annulus gradually collects solid material over long time scales when
the collisional cascade is dormant and \mdotv\ is negligible. Once there is enough material
to collide, episodic cascades generate a large \mdotv\ which is much larger than \mdotz. 

From a large suite of calculations with \r0\ = 0.1--10~km, \mdotz\ = $10^5 - 10^{13}$~\gs,
and $a$ = 0.5--3~\rsun\ (KB2017), the solid mass in the equilibrium state is 
\begin{eqnarray*}
M_{d, eq} & \approx & 1.5 \times 10^{19}~{\rm g} 
\left ( { \dot{M} \over 10^{10}~{\rm g~s^{-1}} } \right )^{1/2}
\left ( { 0.6~\msun \over \mwd } \right )^{9/20}
\left ( { \r0 \over {\rm 1~km} } \right )^{1.04}
\left ( { \rho_s \over {\rm 3~g~cm^{-3}} } \right )^{9/10} \\
 & & ~~~~~~~~~~~~~~~~\left ( { 0.01 \over e } \right )^{4/5}
\left ( { \Delta a } \over {0.2 a} \right )^{1/2}
\left ( { a_0 \over \rsun } \right )^{43/20} ~~~~~~~~~~ \r0 \gtrsim {\rm 1~km} ~ . ~~~~ (5)
\label{eq: meq}
\end{eqnarray*}
\addtocounter{equation}{1}
Aside from the numerical coefficient, the dependence of the equilibrium mass in solids on 
the seven physical variables in eq.~5 is a consequence of simple collision theory (KB2017).
When a system has episodic cascades, the maximum mass is close to the equilibrium mass for 
large \mdotz\ (KB2017).

The expression for $M_{d, eq}$ in eq.~\ref{eq: meq} helps us establish approximate 
conditions for episodic collisional cascades.  In any swarm of solids, the minimum mass in 
solids required for a cascade is $M_{d, min} \approx N_{min} m_0$ with $N_{min} \approx$ 2. 
When $M_{d, min} \ge M_{d, eq}$, the system cannot find an equilibrium and oscillates 
between the high and low states.  Fixing all variables except $a_0$, \mdotz, and \r0\ in 
eq.~\ref{eq: meq} at their nominal values, setting $N_{min}$ = 2 leads to a simple 
estimate for the maximum \mdotz\ in the episodic regime:
\begin{equation}
\dot{M}_{max, ch} \lesssim 3 \times 10^4~{\rm g~s^{-1}} 
\left ( { \r0\ \over {\rm 1~km} } \right )^{3.92} 
\left ( { a_0 \over \rsun } \right )^{-86/20} ~ .
\label{eq: mdotmax} 
\end{equation}
Systems with \mdotz\ $\gtrsim \dot{M}_{max, ch}$ ($\lesssim \dot{M}_{max, ch}$) always
(never) achieve a steady-state with the equilibrium mass in eq.~\ref{eq: meq}.  For our 
annulus with $a_0$ = 1.16~\rsun, the maximum \mdotz\ required for the episodic regime 
ranges from $3 \times 10^4$~\gs\ for \r0\ = 1~km to $8 \times 10^7$~\gs\ for \r0\ = 10~km to 
$7 \times 10^{11}$~\gs\ for \r0\ = 100~km. The full suite of simulations confirms 
this general result.

The two types of collisional cascades generate different evolutionary sequences for 
gaseous disks orbiting the central white dwarf. To illustrate this behavior, we again 
begin with a discussion of simulations with \r0\ = 1~km and various \mdotz. In these
calculations, the smooth time evolution of \mdotv\ leads to a fairly calm gaseous disk 
with a constant accretion rate onto the white dwarf.

\begin{figure} 
\includegraphics[width=6.5in]{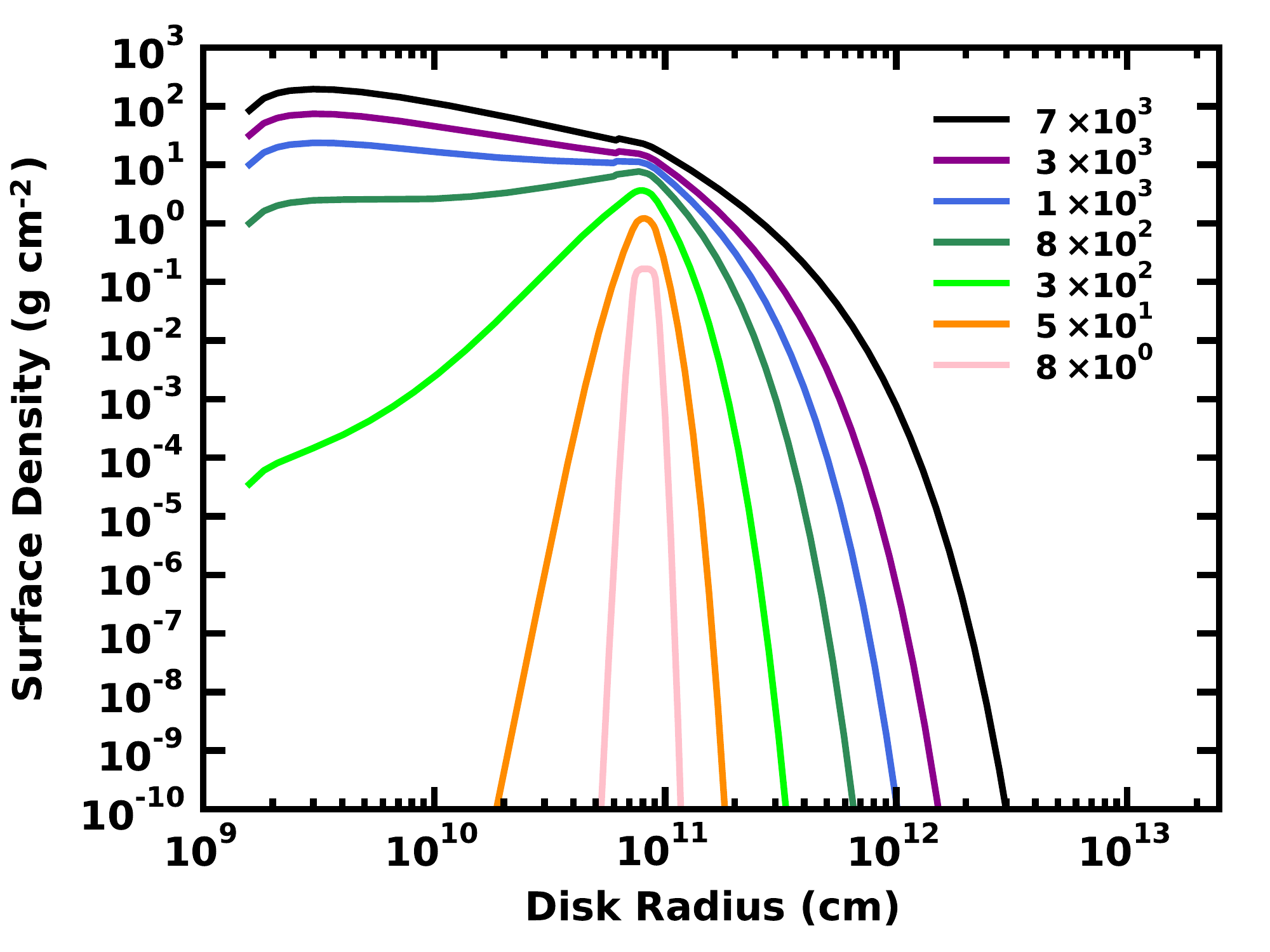}
\vskip 3ex
\caption{
Snapshots of $\Sigma_g(a)$ for a gaseous disk fed by solids from a collisional
cascade with \r0\ = 1~km and \mdotz\ = $10^{13}$~\gs. The legend indicates 
times (in yr) for each snapshot. At 5--10~yr, a gaseous ring is roughly 
centered on the annulus of solids at $a$ = 1.16~\rsun. As the cascade adds
more material to the ring, it expands. After 200--300~yr, material begins to 
reach the surface of the white dwarf. As the surface density grows at the
inner edge of the disk, material expands well beyond the Roche limit and
reaches $a \approx$ 10--20~\rsun\ in $10^4$~yr.
\label{fig: sigma1}
}
\end{figure}

Fig.~\ref{fig: sigma1} illustrates several snapshots of the gas surface density for a 
system with \r0\ = 1~km and \mdotz\ = $10^{13}$~\gs. Initially, the disk contains no
gas; $\Sigma_g$ = 0 for all $a$.  When the cascade begins at $t$ = 8~yr (pink curve), 
vaporized solids generate a narrow torus of gas at $a \approx$ 1.15~\rsun. As vaporization
continues to place gas in this annulus, viscosity spreads the gas to smaller and to larger 
radii \citep[see also][]{lbp1974}. After a few hundred years (lime curve), the gas begins 
to accrete onto the central star. Over the next few hundred years, the surface density of
gas near the white dwarf photosphere grows to rival the surface density of gas at the Roche
limit (dark green and blue curves). By $t$ = 3000--7000~yr, the gaseous disk has evolved 
into a structure where the surface density falls monotonically from 2--3 stellar radii to 
10--30 times the Roche limit. As the evolution continues, the inner disk maintains a roughly 
static structure. Beyond the Roche limit, however, the outer disk radius continues to expand.

\begin{figure} 
\includegraphics[width=6.5in]{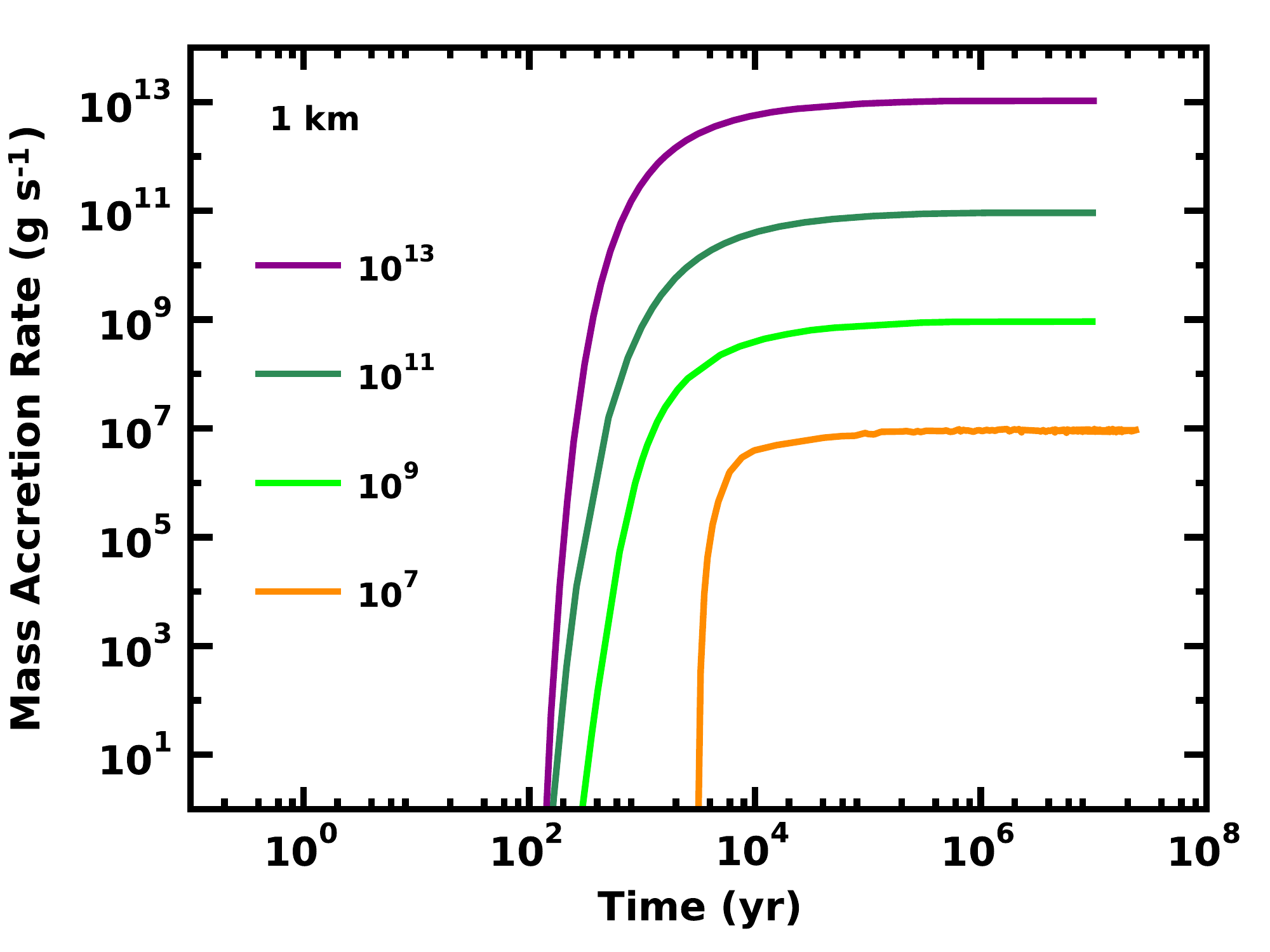}
\vskip 3ex
\caption{
Time evolution of \mdots, the accretion rate onto the central white dwarf, 
during a collisional cascade. Solids with radius \r0\ = 1~km are added to an 
annulus near the Roche limit at rates indicated in the legend. Destructive 
collisions grind the solids into 1~\mum\ particles. Radiation from the central 
star vaporizes these small particles; the resulting gas is added to an annulus 
within a circumstellar disk. Once the collisional cascade develops, the accretion 
rate onto the white dwarf smoothly increases. As the evolution proceeds, the 
accretion rate onto the white dwarf approaches the input rate for solids at 
the Roche limit.
\label{fig: mdot1km}
}
\end{figure}

Fig.~\ref{fig: mdot1km} plots the evolution of \mdots, the mass accretion rate of gas from 
the disk onto the white dwarf for this model (dark magenta curve) and other models with 
\r0\ = 1~km and smaller \mdotz\ (as
listed in the legend). At early times, the collisional cascade is dormant. Once the 
cascade begins, vaporized solids generate a thin annulus of gas. Eventually, viscous 
evolution transports the gas from the Roche limit onto the white dwarf. The derived
\mdots\ then grows dramatically, rising from 1--10~\gs\ to a rate approaching \mdotz\ in 
a few thousand years. After $10^4 - 10^5$~yr, \mdots\ reaches a plateau. In systems 
with large \mdotz\ ($\gtrsim 10^9$~\gs), \mdots\ $\approx$ 0.9\mdotz. When \mdotz\ is 
smaller ($\lesssim 10^9$~\gs), \mdots\ oscillates about this equilibrium rate.

For any \mdotz, the time scale to reach the plateau phase depends on the viscosity. In
our calculations, the viscosity is sensitive to $\alpha$ and to the gas temperature 
$T_g$ (eq.~\ref{eq: disk-nu}).  Systems with factor of ten smaller (larger) $\alpha$ 
reach the plateau phase ten times more slowly (rapidly). Similarly, raising (lowering)
$T_g$ by a factor of two decreases (increases) the viscous time by a factor of two. 

Although the time scale to reach the plateau phase depends on $\alpha$ and $T_g$, the 
plateau \mdots\ is insensitive to either variable. In all of our calculations with 
\r0\ = 1~km, the plateau rate is always roughly 90\% of the input \mdotz. 

Based on a broad suite of simulations with \r0\ = 0.1--10~km and a variety of \mdotz,
the mass in the disk is independent of \r0, $\rho_s$, and other properties of the solids.
In addition to \mdotz, the mass in gas depends on $\alpha$ and $T_g$:
\begin{equation}
M_g \approx 2.2 - 2.4 \times 10^{22}~{\rm g} 
\left ( { \dot{M} \over 10^{10}~{\rm g~s^{-1}} } \right )
\left ( { {\rm 1500~K} \over T_0 } \right )
\left ( { 10^{-3} \over \alpha } \right ) ~ .
\label{eq: mgas}
\end{equation}
Comparing this numerical result with eq.~\ref{eq: meq}, the equilibrium mass in the gaseous 
disk is roughly three orders of magnitude larger than the equilibrium mass in solids.

\begin{figure} 
\includegraphics[width=6.5in]{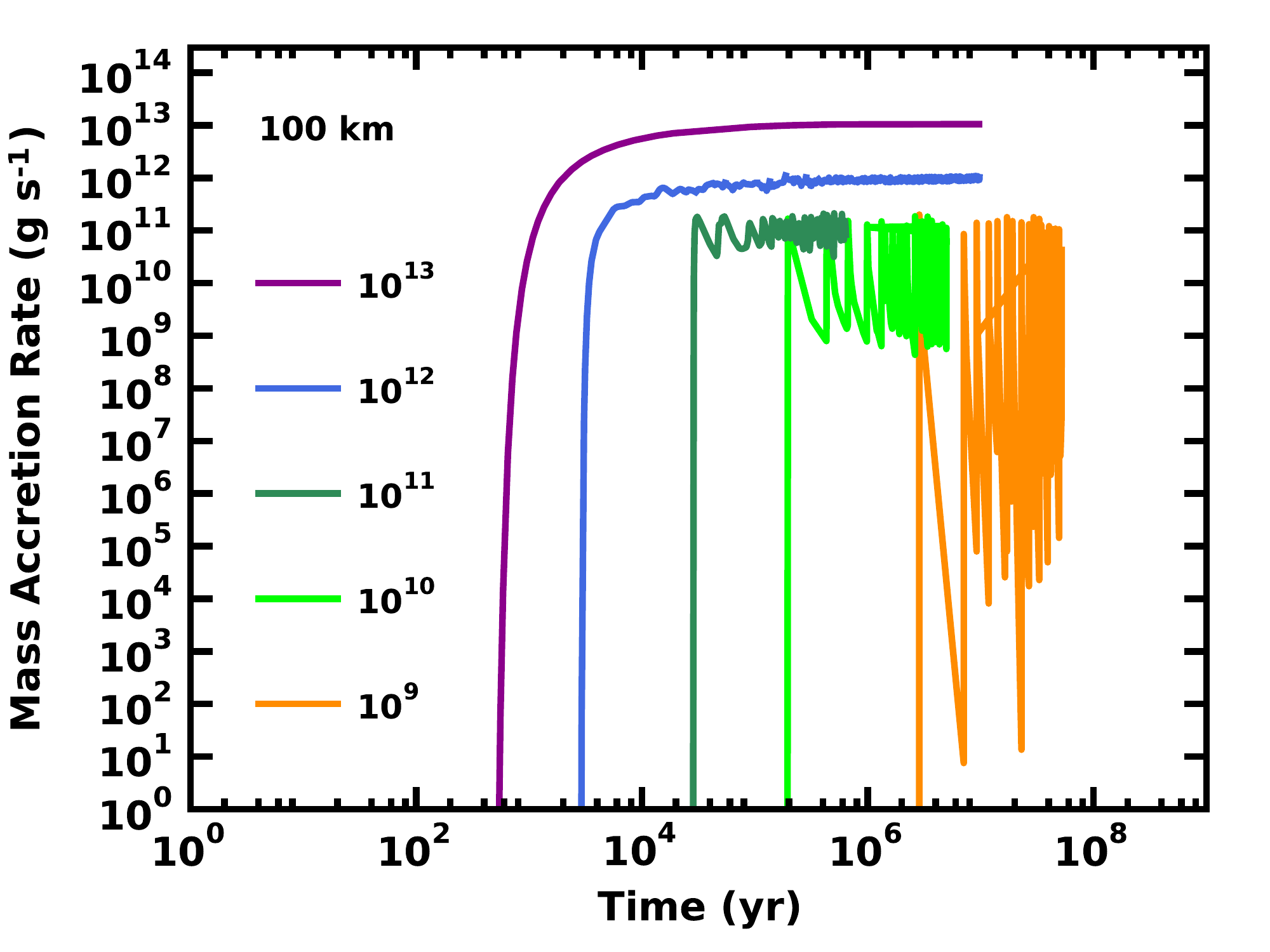}
\vskip 3ex
\caption{
As in Fig.~\ref{fig: mdot1km} for \r0\ = 100~km. For large input rates of 
solids, $\mdotz \gtrsim 3 \times 10^{11}$~\gs, the accretion rate onto the
white dwarf is fairly constant in time. At smaller rates, the accretion 
rate onto the white dwarf varies dramatically.
}
\label{fig: mdot100km}
\end{figure}

In calculations with larger \r0, the structure of the disk and the rate of mass accretion 
onto the white dwarf oscillate between low and high states. Fig.~\ref{fig: mdot100km} shows 
the evolution for five simulations with \r0\ = 100~km and \mdotz\ = $10^9 - 10^{13}$~\gs. 
When \mdotz\ is large, the evolution is smooth. The mass accretion rate onto the white dwarf 
gradually increases to the plateau value in $10^4 - 10^5$~yr.  As the input \mdotz\ drops, 
high \mdots\ states becomes more and more episodic. On time scales of 0.1--1~Myr, the 
variations in \mdots\ grow from $\pm$5\% at \mdotz\ = $10^{12}$~\gs\ to $\pm$50\% at 
\mdotz\ = $10^{11}$~\gs\ to $\pm$2 orders of magnitude at \mdotz\ = $10^{10}$~\gs\ to 
$\pm$6--8 orders of magnitude at \mdotz\ = $10^{9}$~\gs. 

Although the time scale for changes in \mdots\ depends on $\alpha$ and $T_g$, the overall 
fluctuations are sensitive only to \r0\ and \mdotz. When \r0\ is large and \mdotz\ is small, 
\mdots\ varies by many orders of magnitude over 0.1--1~Myr time scales. Smaller \r0\ and 
larger \mdotz\ smooth out the variations in \mdots.

For calculations with \r0\ = 30--100~km and \mdotz\ $\gtrsim 10^{12}$~\gs, the typical 
mass in the disk is close to the equilibrium mass in eq.~\ref{eq: mgas}. Due to stochastic 
variations in \mdotv, the mass varies by $\pm$10\%--20\% about the equilibrium mass. 
Smaller \mdotz\ leads to large oscillations in \mdotv\ and similarly large variations 
in $M_g$.  When the collisional cascade generates large \mdotv, the \textit{maximum} disk 
mass is roughly $10^{23}$~g for $T_0$ = 1500~K and $\alpha$ = $10^{-3}$.  This maximum 
mass scales with $\alpha$ and $T_0$ as in eq.~\ref{eq: mgas}.  During low states, 
$M_g \lesssim 10^{10}$~g. Throughout a single oscillation, the total variation in 
$M_g$ is more than fifteen orders of magnitude.

For all systems, the fraction of time in the high state is a strong function of \r0\ and 
\mdotz\ (KB2017). In principle, the variation of the mass in gas provides a reasonable 
proxy for the time spent in high and low states. However, detecting the gas depends on the 
thermodynamic state of the disk and the ability of spectrographs to distinguish absorption 
or emission lines produced in the disk from those generated in the white dwarf photosphere. 
The X-ray luminosity probably enables a clearer picture. Despite uncertainties in the likely
X-ray temperature, the X-ray luminosity provides an instantaneous measure of the accretion 
rate from the disk onto the white dwarf \citep[e.g.,][and references therein]{kuulkers2006,
pretorius2012,reis2013}.

To predict the X-ray luminosity, we assume that half of the accretion energy is radiated in a
boundary layer between the disk and the white dwarf \citep[e.g.,][]{lbp1974} or at the base 
of a magnetic accretion column \citep[e.g.,][]{ghosh1979}. Ignoring factors of order unity, 
$L_X \approx G M_\star \mdots\ / 2 R_\star$.  The fraction of time with 
$L_X / L_\odot \gtrsim 10^{-n}$ with integer $n$ follows directly from our calculations.

\begin{figure} 
\includegraphics[width=6.5in]{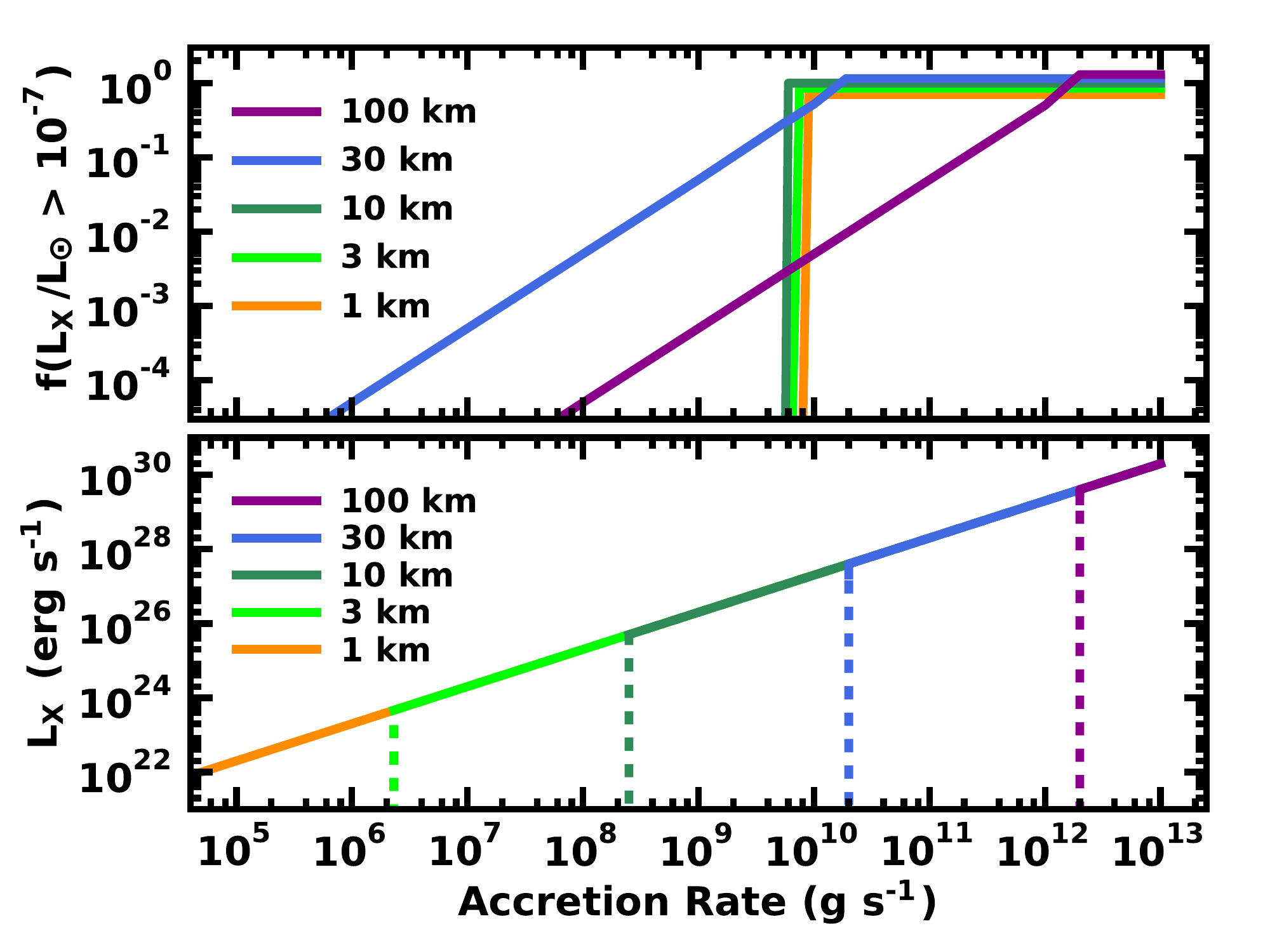}
\vskip 3ex
\caption{
Predictions for the X-ray luminosity $L_X$ of metallic line white dwarfs accreting
from a viscous disk fed by a collisional cascade.
\textit{Lower panel}: $L_X$ as a function of \mdotz\ for \r0\ = 1--100~km as 
indicated in the legend. Solid (dashed) lines indicate regimes where the 
collisional cascade and accretion onto the white dwarf are steady (intermittent).
\textit{Upper panel}: Predicted fraction of time accreting white dwarfs spend 
with $L_X / L_\odot > 10^{-7}$ as a function of \mdotz\ and \r0. For clarity, 
some predicted values have been displaced vertically or horizontally.  Systems 
with \r0\ $\lesssim$ 10~km have no episodic evolution and spend all of their 
time with a fraction of zero or unity. Episodic evolution in systems with
$\r0\ \gtrsim$ 30~km produces a broader range of accretion luminosities for an
input \mdotz.
}
\label{fig: xray}
\end{figure}

Fig.~\ref{fig: xray} plots our results. In the lower panel, the solid curve indicates the 
X-ray luminosity as a function of the accretion rate onto the white dwarf, 
$L_X \approx 5 \times 10^{16} \mdots$~erg~s$^{-1}$.  When the mass input rate for solids 
is large, 
\mdotz\ $\gtrsim \dot{M}_{0, crit}$ $\approx$ $3 \times 10^4 ~ (r_0 / {\rm 1~km})^{3.92}$~\gs,
the collisional cascade steadily feeds the gaseous disk. Accretion onto the white dwarf 
is also steady. When \mdotz\ falls below $\dot{M}_{0, crit}$, the X-ray luminosity varies 
between low states with $L_X \approx$ 0 and $L_X \approx L_{X, crit}$ 
$ \approx 5 \times 10^{16} \dot{M}_{0, crit}$~erg~s$^{-1}$. The vertical dashed lines in 
the lower panel indicate the critical \mdotz\ for values of \r0\ listed in the legend. 

The upper panel in Fig.~\ref{fig: xray} shows predictions for the fraction of time a system 
spends with $L_X / L_\odot \gtrsim 10^{-7}$.  For systems with \r0\ $\lesssim$ 10~km, 
\mdots\ and the X-ray luminosity are nearly constant in time. The white dwarf is above 
the reference luminosity all of the time or none of the time. When \r0\ is larger,
episodic evolution of \mdots\ leads to a broad range of X-ray luminosities for any
combination of \r0\ and \mdotz. In our calculations, swarms with \r0\ = 30~km are 
detectable X-ray sources at least some of the time. The windows for detecting swarms
with \r0\ $\gtrsim$ 100~km are much smaller.

As an illustration of the utility of Fig.~\ref{fig: xray}, we consider two examples.
For a system with \r0\ = 10~km, the critical accretion rate and X-ray luminosity are 
$\dot{M}_{0, crit} \approx 2.5 \times 10^{8}$~\gs\ and 
$L_{X, crit} \approx 10^{25}$~erg~s$^{-1}$. When 
\mdotz\ $\gtrsim 2.5 \times 10^{8}$~\gs, we expect an accretion luminosity along the
diagonal solid curve in the lower panel. For smaller accretion rates, the system has
$L_X$ between zero and $L_{X, crit}$. This critical $L_X$ is smaller than $10^{-7}$~\lsun.
Thus, the fraction of time with $L_X / L_\odot > 10^{-7}$ is zero.

For a second example, suppose a metallic line white dwarf has $L_X \approx 10^{-7}~\lsun$.  
The implied accretion rate is $7.6 \times 10^9$~\gs.  In the lower panel, this rate is 
larger than (smaller than) critical rates for \r0\ = 1--20~km (25--100~km). Thus, this 
system could have a steady-state collisional cascade with $\r0\ \lesssim$ 10~km or an 
intermittent cascade with $\r0 \gtrsim$ 25~km. Moving to the upper panel, the fraction of
time systems with \r0\ = 1--10~km spend in this high $L_X$ state is either unity (if
\mdotz\ $\gtrsim 7.6 \times 10^9$~\gs) or zero (if \mdotz\ $\lesssim 7.6 \times 10^9$~\gs).
Systems with \r0\ = 30~km (100~km) spend all of their time at or above this $L_X$ when
\mdotz\ $\gtrsim 2 \times 10^{10}$~\gs\ (\mdotz\ $\gtrsim 2 \times 10^{12}$~\gs). 
Otherwise, the fraction of time spent at this $L_X$ is roughly $\mdotz / \dot{M}_{0, crit}$. 

In these examples, a single $L_X$ observation places few constraints on the model.  Surveys 
with at least 10--20 detections or upper limits begin to carve out allowed sections of
(\r0, \mdotz) space. Combined with estimates of accretion rates derived from photospheric
abundance measurements, these data begin to test the model.

\section{DISCUSSION}
\label{sec: disc}

\subsection{Simple Physical Model for Metallic Line White Dwarfs}
\label{sec: disc-mod}

In KB2017 and this paper, we have described a simple physical model for the delivery 
of metals to a white dwarf photosphere. In this model, at least one mechanism places
large asteroids with radius \r0\ and mildly eccentric orbits into a ring near the 
Roche limit of a white dwarf at a constant rate \mdotz.  Interactions among the solids 
generate a collisional cascade, which produces swarms of 1~\mum\ dust grains. 
Vaporization of the grains feeds a ring of gas.  Viscous processes spread the ring 
into a disk; material from the disk accretes onto the central white dwarf.

The main parameters in this picture are \r0\ and \mdotz. When \mdotz\ $\gtrsim$ 
$\dot{M}_{0, crit}$ $\approx$ $3 \times 10^4 ~ ({\rm r_0 / 1~km})^{3.92}$~\gs,
the solids find an equilibrium state where 
(i) \mdotv, the vaporization rate of the grains, roughly equals \mdotz\ and 
(ii) the mass of the gaseous disk and the X-ray luminosity are roughly constant in time. 
Small solids with \r0\ $\lesssim$ 1~km find equilibrium with relatively small \mdotz. 
Larger solids with \r0\ = 10~km (100~km) require input rates 
\mdotz\ $\gtrsim 2.5 \times 10^8$~\gs\ (\mdotz\ $\gtrsim 2 \times 10^{12}$~\gs).
Once the system achieves equilibrium, metals accrete onto the white dwarf at a rate 
\mdots\ $\approx$ 0.9~\mdotz. Although most of the vaporized metals accrete onto the 
white dwarf, viscosity spreads some of the mass to large distances from the central star.

When \mdotz\ $\lesssim \dot{M}_{0, crit} $, the time for the ring to accumulate enough 
solids for a collisional cascade is longer than the time scale for the cascade to deplete 
the ring of solids. In these circumstances, the system oscillates between low states 
(where the mass in solids slowly grows with time, \mdotv\ $\ll$ \mdotz, and the X-ray 
luminosity is negligible) and high states (where the mass rapidly declines with time, 
\mdotv\ $\gg$ \mdotz, and the X-ray luminosity is substantial).  As long as the viscous 
time scale is shorter than the cycle time between high and low states, the mass accretion 
rate onto the white dwarf varies by orders of magnitude on time scales of thousands to 
millions of years.

In a more realistic system where \mdotz\ and \r0\ vary on short time scales, the behavior 
of the solids and the gaseous disk depends on the variability time scale. When the cascade 
is active, the collision time for the largest objects ranges from the orbital period (for
large \mdotz) to a few months (for small \mdotz).  On this time scale, the cascade adjusts 
the vaporization rate and the production rate of small particles to match the time-variable 
\mdotz\ and \r0. In this way, changes in \mdotz\ and \r0\ can generate substantial variations 
in the IR excess from 1~\mum\ particles and the amount of gas near the Roche limit on week 
or longer time scales.

When $\alpha \approx 10^{-3}$, the viscous time is roughly $10^4 - 10^5$ orbital periods,
$\sim$ 10~yr. In an active cascade , variations in \mdotz\ and \r0\ for the solids produce 
similar scale fluctuations in \mdots. Although variability in $L_X$ correlates with changes 
to the IR excess and the amount of gas near the Roche limit, there is a 10~yr or longer delay 
between rises/drops in $L_X$ and the IR excess. This delay is roughly proportional to $\alpha$.
When the cascade is not active, \mdotv\ and \mdots\ are close to zero. Substantial 
variations in \mdotz\ and \r0\ are invisible. 

In any of the examples we studied, the vertical scale height of the solids remains large.
If solids have $e_0$ = 0.01 and inclination $\imath_0 = e_0 / 2$, the cascade removes 
solids faster than collisional damping can reduce the vertical scale height (KB2017). 
Thus, the solids do not evolve into the optically thick, vertically thin structure 
originally suggested by \citet{jura2003}.

\subsection{Comparisons with Previous Results}
\label{sec: disc-comp}

Although our approach is the first to combine collisional evolution of solids with 
viscous diffusion of gas, other investigators have treated aspects of these phenomena 
in the context of metallic line white dwarfs. In this sub-section, we compare the
methodologies and results of these studies with our own.

\citet{brown2017} consider the tidal destruction, sublimation, and ultimate fate of rocky 
and icy asteroids with periastron distances $q \approx \rstar$. In this situation, large
solids tidally fragment; small solids are tidally stable but sublimate. Delivery of material 
onto the white dwarf then depends on the initial radius, composition, and $q$ for each 
asteroid.  In a manner similar to meteors encountering the Earth, larger fragments ablate 
in the white dwarf atmosphere; icy fragments ablate more rapidly than rocky ones. Smaller 
fragments often sublimate before reaching the atmosphere; the resulting gas then rains down 
onto the white dwarf. With no calculation for the long-term evolution of gas, estimates for 
\mdots\ rely on the rate of direct collisions of asteroid with the white dwarf or assumptions
on the subsequent evolution of gas orbiting the white dwarf.

Our somewhat different analysis of tidal forces (KB2017) indicates that solids with 
\r0\ $\lesssim$ 100~km and $\rho_s$ = 3~\gcmc\ are stable at distances $a \gtrsim 0.7 a_R$, 
where $a_R$ is the distance of the Roche limit from the center of the white dwarf 
\citep[see also][]{veras2017}.  Our calculations do not address the fate of solids at 
smaller $a$. Although our conclusions on the sublimation of large particles are similar
to \citet{brown2017}, the sublimation time for particles with radii $r \gtrsim$ 1~\mum\ is 
longer than the collision time throughout the collisional cascade. Thus, it is reasonable 
for us to neglect sublimation for large particles and focus on the sublimation of the 
smallest particles that feed the gaseous disk. Compared to \citet{brown2017}, our estimates 
for \mdots\ rely on viscous diffusion through a gaseous disk instead of direct collision 
with the white dwarf.

In their investigation of the evolution of disks containing cm-sized solids interacting 
with a viscous, gaseous disk, \citet{metzger2012} solve a radial diffusion equation similar 
to our eq.~\ref{eq: disk-evol} \citep[see also][]{rafikov2011a,rafikov2011b,bochkarev2011,
rafikov2012}. Adopting a very small vertical scale height for the solids, they treat the
back-reaction of the solids on the gas; in our treatment, the much larger vertical scale
height of the solids implies a negligible back-reaction which is safely ignored. While we
adopt a constant $\alpha$ and allow the disk temperature to vary with distance from the
central star, \citet{metzger2012} adopt a constant disk temperature and let $\alpha$ vary 
with radius. For a similar $\Sigma (a)$, the magnitude and variation of $\nu(a)$ -- which 
controls the evolution of the surface density of the gas -- is similar in the two approaches. 
The time scale for a ring of gas to accrete onto the star from some distance $a$ is also similar.

Adopting a fixed initial mass for the solids, $M_d$, \citet{metzger2012} derive \mdots\ as
a function of $M_d$ and other properties of the solids and the gas. When the solids are
optically thin (thick), \mdots\ is roughly 100 times smaller (larger) than the rate generated 
by PR drag, $\mdots \approx 10^{-8}$~\gs.  Interactions between opaque disks of solids and 
gas can also generate large time-variations in \mdots.  In contrast, the \mdots\ derived in 
our calculation depends on an adopted input rate \mdotz\ and typical radius \r0\ for solids 
near the Roche limit; oscillations in \mdots\ occur when \mdotz\ is small and \r0\ is large.

\citet{bear2013} combine an analysis of tidal disruption \citep[similar to][]{brown2017} 
with a simple treatment for viscous disk evolution \citep[compared to][]{metzger2012} to
propose that infalling asteroids drive transient accretion events similar to those associated 
with supermassive black holes at the centers of galaxies \citep[e.g.,][and references 
therein]{rees1988,cannizzo1990,gezari2009,lodato2011,bromley2012,kochanek2016}. Assuming that 
(i) a massive ($\gtrsim 10^{20}$~g) asteroid is completely converted into gas by collisions and 
sublimation, (ii) the gas has a high temperature derived from the kinetic energy of infall, and 
(iii) the gas lies in a vertically thin, optically thick disk, they infer a peak accretion rate 
exceeding $10^{13}$~\gs\ and X-ray luminosity exceeding $10^{30}$~\ergs\ over a typical lifetime 
of a few days to a few weeks.

In our approach, the collision velocities of asteroid fragments on $e \approx$ 1 orbits are set
by the velocity dispersion of the fragments, not their orbital velocity. Numerical simulations 
suggest tidal disruption of asteroids generates a long string of fragments along an orbit with 
similar $e$ and $\imath$ as the original asteroid \citep[e.g.,][]{debes2012a}. Unless other 
processes change $e$ and $\imath$, we expect low velocity collisions among the fragments to produce 
a cascade similar to those calculated here. With most interactions near periastron of a large 
$e$ orbit, modest vaporization results in gas ejected from the orbit. Subsequent evolution
of the gas depends on the collision frequency, the vaporization of small particles, and the temperature 
and viscosity of the gas. Detailed numerical simulations are necessary to learn the fate of this
material; we speculate that the white dwarf accretes gas with a typical temperature of 
$10^4 - 10^5$~K at some modest background rate set by continuous vaporization of small particles, 
with occasional flares from the production of debris from collisions of larger fragments.

To constrain the frequency and sizes of accreted asteroids among metallic line white dwarfs, 
\citet{wyatt2014} explore analytical estimates and Monte Carlo calculations which consider 
how derived accretion rates depend on the gravitational settling time in the white dwarf atmosphere, 
the distributions of accretion rates and masses for accreted asteroids, and the typical time scale 
for vaporized solids to accrete onto the white dwarf. Their results suggest that white dwarfs 
accrete solids ranging in size from $\lesssim$ 1--10~km to 100--1000~km at rates ranging from 
$ \sim 10^6$~\gs\ to $\sim 10^{11}$~\gs\ with a median of $10^{-8}$~\gs. Smaller solids are 
much more common than larger solids.  In DA white dwarfs with short settling times, accretion of 
1--30~km objects is nearly continuous. Asteroids with radii of 30--100~km dominate the pollution 
of non-DA white dwarfs with much longer settling times. The time scale for a gaseous disk to 
deposit metals onto the white dwarf photosphere is 20--1000~yr. 

Our calculations are consistent with these results. Disk time scales of 20--1000~yr imply 
$\alpha$ = 0.05--$10^{-3}$, close to the range deduced in the accreting white dwarfs of cataclysmic 
variables \citep[e.g.,][]{smak1999,king2007,kotko2012}. For the range of accretion rates derived 
in \citet{wyatt2014}, our analysis suggests continuous accretion for \r0\ = 1--10~km and episodic 
accretion for \r0\ $\gtrsim$ 30~km. The division between episodic and continuous accretion agrees 
rather well with the expectations for DA and non-DA white dwarfs from \citet{wyatt2014}. 

Overall, these and other analyses paint a fairly coherent picture for the transport of metals 
from a region near the Roche limit onto the surface of a white dwarf. In some fashion, dynamical 
processes regularly transport material from large $a$ to the Roche limit of the white dwarf 
\citep{wyatt2014}.  Well inside the Roche limit, tidal forces disrupt the solids into fragments 
\citep[e.g.,][]{holsapple2006,holsapple2008,debes2012a,bear2013,veras2014a,veras2017,brown2017}. 
If large solids or disrupted fragments begin to collide, they are rapidly ground into small dust 
grains (KB2017, this paper).  UV radiation from the white dwarf rapidly sublimates small solids 
and slowly evaporates larger objects \citep[this paper; see also][]{veras2015d,brown2017}. 
Viscous processes transport vaporized solids to the central star \citep[this paper; see
also][]{metzger2012,bear2013}. 

Despite the attractiveness of this picture, there are many uncertainties. Plausible mechanisms 
for delivering solids to the Roche limit have few observational constraints 
\citep[e.g.,][]{debes2012a,frewen2014,veras2015b,payne2016,hamers2016,antoniadou2016,brown2017}. 
Models for disk evolution are rather simple, with limited treatment of interactions between the
solids and the gas. Aside from explaining the transport of solids to the white dwarf, it is not 
yet clear whether the model can account for other aspects of observations. In the next few
sub-sections, we comment on these issues in more detail.

\subsection{Delivery of Solids to the Roche Limit}
\label{sec: disc-del}

Various groups have considered the delivery of solid objects to a volume within the 
Roche limit of a white dwarf \citep[e.g.,][]{debes2002,veras2014c,veras2015a,bonsor2015,
antoniadou2016,veras2016,payne2017,brown2017,petrovich2017,stephan2017,caiazzo2017}. 
All models begin with
a main sequence star and a surrounding planetary system. While on the main sequence,
the nearly constant luminosity of the central star establishes the `snow line', which 
marks the boundary between an inner `terrestrial' region with little volatile material
and an outer `icy' region where volatiles can condense from the gas phase onto solids
\citep[e.g.,][]{kenn2008a}. 
As the central star evolves into a red giant and then an asymptotic branch giant, the 
increasing luminosity of the central star moves the snow line outward and bakes solids
between the `original' and `new' snow lines \citep{stern1990,parriott1998,villaver2007,
dong2010,bonsor2011,veras2013,mustill2014,malamud2016,malamud2017a,malamud2017b}. 
The slowly decreasing mass of the central star also results in an expansion and possible 
destabilization of orbiting solids. If destabilization is sufficiently traumatic, solids 
can attain extremely eccentric orbits with periastra inside the Roche limit of the white 
dwarf central star. After some number of passes close to the white dwarf, tidal forces 
disrupt the solids into much smaller objects.

In addition to a lack of agreement on the mechanism(s) that place(s) solids on very 
eccentric orbits, it is unclear how the orbits of small solids objects evolve from 
$e$ = 0.99 to $e \lesssim$ 0.1. Although we do not address this issue directly, our
calculations provide some constraints on likely outcomes of plausible paths from high
$e$ to small $e$ orbits. In mechanisms where long-term dynamical processes such as 
PR drag gradually circularize the orbits \citep[e.g.,][]{veras2014a}, the long-term 
outcome is probably similar to that outlined here (see also KB2017): once the orbits 
of the solids cross, a collisional cascade generates small objects which vaporize and 
feed a gaseous disk. Although the geometry and observable properties of the cascading 
solids and the disk depend on the mass and orbital properties of the incoming solids, 
outcomes are probably similar to those outlined here.

If long-term dynamical solutions are unreliable, our results indicate that collisions
are insufficient to reduce $e$ and $i$ dramatically on their own. Some other mechanism
is required. If two asteroids with large $e$ collide inside the Roche limit, they
probably vaporize. The gas released from this collision might be sufficient to reduce
$e$ for other asteroids following similar paths around the white dwarf.  
Interactions between incoming asteroids with the magnetic field of the white dwarf 
may generate sufficient electromagnetic induction and Ohmic dissipation to reduce $e$
on time scales ranging from a few Myr to several Gyr \citep[Bromley \& Kenyon, in 
preparation; see also][]{li1998}.

Incorporating any mechanism for the delivery of solids to the Roche limit into our calculations 
requires more comprehensive theoretical predictions of outcomes for the evolution of gas
and solids as the central star begins to evolve into a white dwarf.  For example, detailed 
predictions for the distributions of $\Sigma$, $a$, $e$, and $\imath$ for swarms of solids 
outside the Roche limit would allow us to learn how outcomes for the delivery of solids to 
the white dwarf depend on the delivery mechanism.

\subsection{Improved Models for Evolution of the Gas}
\label{sec: disc-gas}

In this first exploration of the evolution of a gaseous disk fed by a collisional cascade, 
we assume the gas is axisymmetric and adopt a simple prescription for the disk temperature
\citep[see also][]{jura2008,melis2010b}.  Although our approach is reasonable, we outline 
several possible improvements for future studies. 

By analogy with the circumstellar disks in cataclysmic variables \citep{meyer1982,mineshige1983,
cannizzo1984} and pre-main sequence stars \citep{daless1998,najita2011,najita2013}, we expect 
a complex temperature structure for gas fed by a collisional cascade. For optically thin disks
where the solids have negligible vertical scale height ($H \ll H_g$), \citet{melis2010b} derived 
the radial temperature structure outside 25~\rstar\ for an ionized gas with Mg, Si, Ca, and Fe.  
However, our disks have $H > H_g$; an improved calculation should include interactions between 
the gas and dust. As one example, combining the formalism of \citet{najita2013} with our derived 
radial surface density distribution should enable calculations of the radial and vertical 
temperature structure from the inner edge of the disk out past the Roche limit. Since the 
evolution of $\Sigma$ depends on $T_g$, a better prescription for $T_g$ would yield a better 
connection between the input (\r0, \mdotz) and the output \mdots\ and $L_X$.

Our solution of the radial diffusion equation assumes the white dwarf has a negligible magnetic
field. When the field has a modest strength, it can truncate the disk and channel gas directly
onto the white dwarf photosphere \citep[e.g.,][and references therein]{ghosh1979,metzger2012,
mukai2017}. In some configurations, the field might be able to trap dust and gas in the 
magnetosphere \citep[e.g.,][]{farihi2017a}. If most metallic line white dwarfs have modest to
large magnetic fields, it will be necessary to address whether the field has any impact on 
small particles in the collisional cascade.

Eventually, it should be possible to relax the assumption of an axisymmetric disk. Our 
calculations assume that vaporization places material uniformly within an annulus. In a 
real system, vapor generated from collisions of 1~km and larger solids is clumpy. Naively, 
we expect more clumpiness from collisions of larger objects. The azimuthal structure of 
the disk then depends on the viscous time scale relative to the time scale for collisions
of the largest objects. When the viscous time scale is shorter than the collision time,
viscosity rapidly spreads out blobs of gas; the disk is more axisymmetric. When the 
collision time is shorter than the viscous time, small particles are vaporized faster than
viscous shear can spread the gas. The disk is then more asymmetric.

\subsection{Gas Drag}
\label{sec: disc-drag}

Although we have not included interactions between the solids and gas in our simulations,
it is straightforward to show that the gas has little impact on collisional cascades with
$e_0$ = 0.01. In protoplanetary and circumplanetary disks, massive solid objects modify 
the density structure of the gas \citep[e.g.,][and references therein]{ward1997,tanaka2002,
masset2003,nelson2004,ida2008a,lyra2010,bk2011b,bk2013}; gravitational torques generated 
by these structures induce radial migration of the solids. For gaseous disks inside the 
Roche limit, the time scale for migration is rarely shorter than the viscous time but is
much shorter than the cooling time for the white dwarf. Thus, stable solids \textit{could} 
migrate on interesting time scales.  However, these time scales are much longer than the 
collision time. In the context of our calculations, migration is safely ignored.

In any disk with finite pressure, the gas orbits the central star somewhat more slowly than
solids following Keplerian orbits \citep[e.g.,][and references therein]{ada1976,weiden1977a,
youdin2004a,youdin2010,youdin2013}. The solids then feel a headwind, $\Delta v = v_g - v_K$,
where $v_g$ is the orbital velocity of the gas. The
response to the headwind depends on the `stopping time' $t_s$ required for an orbiting 
particle to encounter its mass in gas. When the stopping time is much longer than the orbital 
period, $T$, solids do not respond to the gas. Solids with $t_s \ll T$ are entrained in the 
gas and drift inward or outward on the local viscous time. In between these two limits, solids 
drift radially inward with a maximum drift velocity $\Delta v$ \citep{ada1976,weiden1977a}.

To derive the radial drift speed as a function of particle radius, we follow the
formalism of \citet{weiden1977a} and consider drift in axisymmetric disks with a
radial surface density generated by our solution to the diffusion equation, 
eq.~\ref{eq: disk-evol}.  Given a derived $\Sigma_g(a, t)$ and an adopted $\gamma$, 
$\mu$, and $T_g(a)$, we infer the gas density $\rho_g$ and pressure $P_g$ required 
to establish radial drift rates. Defining $g$ = $v_K^2 / a$ as the central gravity and 
$\delta g = \rho_g^{-1} \partial P_g / \partial R$ as the residual gravity, the 
headwind is \citep{weiden1977a}
\begin{equation}
\Delta v \simeq - \left ( {\delta g \over {2 g} } \right ) v_k ~ . 
\label{eq: deltav} 
\end{equation}
As a measure of the drift speed relative to the local orbital velocity, it is useful
to define 
\begin{equation}
\eta = {\Delta v \over v_K} \simeq - \left ( { \delta g \over {2 g} } \right ) ~ .
\label{eq: eta}
\end{equation}

\citet{weiden1977a} decomposed the motion into a radial drift $u$ and a transverse drift 
$w$, which depend on $\Delta v$ and particle size.  \citet{take2001} later generalized 
the \citet{weiden1977a} approach to disks where Poynting-Robertson (PR) drag and radiation 
pressure are relevant. As in \citet{knb2016}, we derive $u$ and $w$ as a function of particle 
size for a system with radiation pressure but no PR drag. As discussed in KB2017, the time
scale for PR drag is much longer than the collision time. Our goal is to learn
whether gas drag or radiation pressure can produce radial drift of small particles 
on time scales shorter than the collision time. Thus, PR drag is safely ignored.

\begin{figure} 
\includegraphics[width=6.5in]{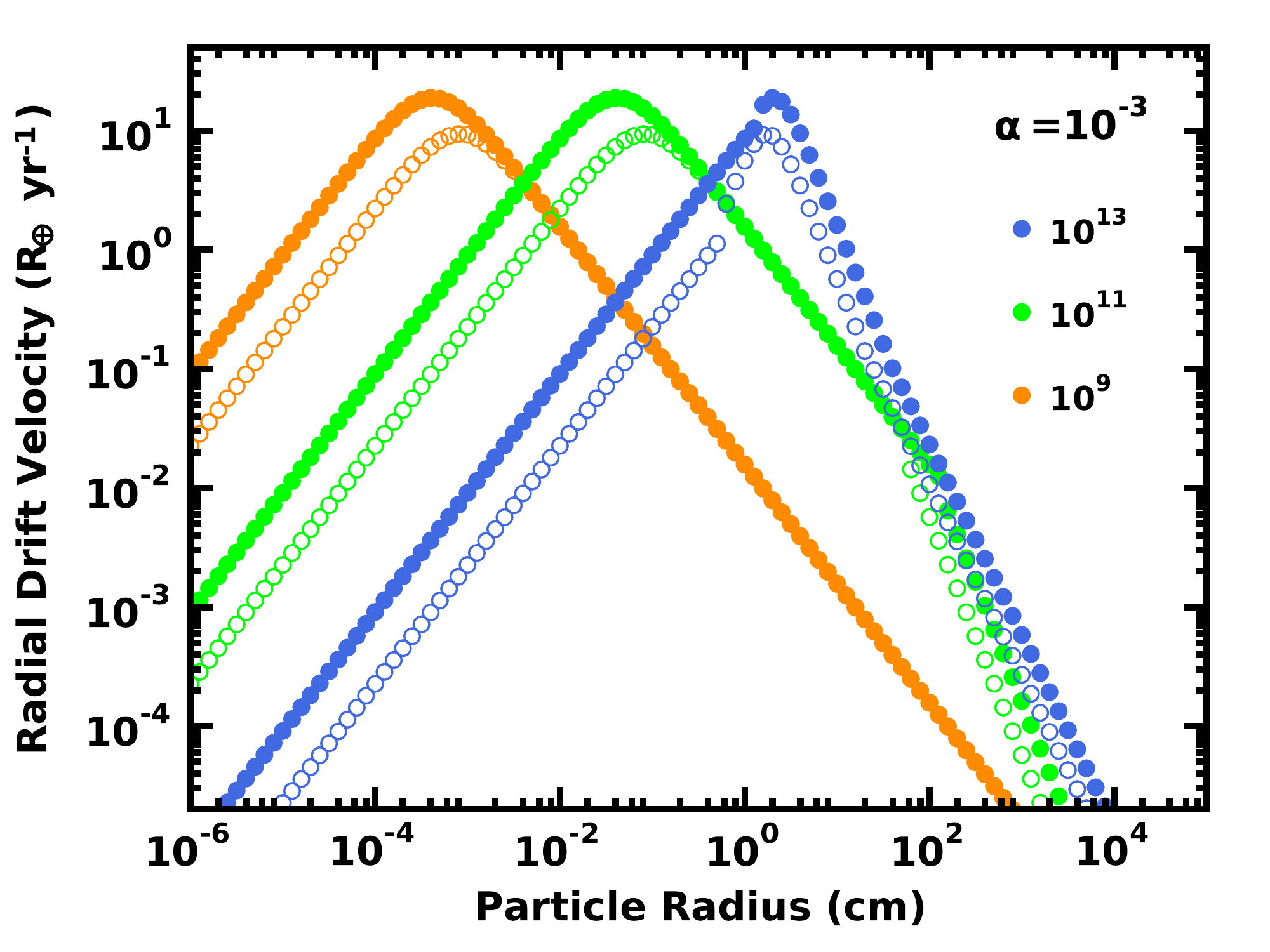}
\vskip 3ex
\caption{
Radial drift velocity as a function of particle size for solids at $a_0$ = 
1.15~\rsun\ within a steady-state gaseous disk generated by vaporization of 
1~\mum\ and smaller particles. The legend indicates the input \mdotz\ in solids 
and the $\alpha$ for the gas. Particles in disks with $T_0$ = 1500~K (open circles)
have somewhat smaller radial drift velocity than those in disks with $T_0$ = 3000~K 
(filled circles).  At the maximum radial drift velocity of 5 (10) $R_\oplus$~yr$^{-1}$
for $T_0$ = 1500~K ($T_0$ = 3000~K), the collisional cascade destroys small solid 
particles before radial drift removes them from the annulus.
}
\label{fig: drag1}
\end{figure}

Fig.~\ref{fig: drag1} illustrates results for particle drift at 
$a$ = 1.15~\rsun\ (within the ring of solids) in a gaseous disk with the equilibrium mass,
$\alpha = 10^{-3}$, no radiation pressure, and $T_0$ = 1500~K (open circles) 
or $T_0$ = 3000~K (open circles).  The legend indicates the input \mdotz. 
For all systems, the maximum drift rate relative to the gas is 
10--20~\rearth\ yr$^{-1}$.  With $u \propto T_g$ \citep{weiden1977a}, 
particles in cooler disks drift more slowly.  Moving away from each maximum 
in the Figure, smaller particles have successively lower drift rates until 
they orbit with the gas. Larger particles drift more slowly because they have 
too much inertia for the gas to move. These particles follow Keplerian orbits. 

Overall, it is clear from Fig.~\ref{fig: drag1} that the radius of particles with
the maximum drift velocity becomes smaller with smaller input \mdotz. Lower input
rates generate disks with lower surface density $\Sigma_g$ and mass density $\rho_g$.
During one orbit of the central star, particles see less mass in disks with smaller 
\mdotz\ than in disks with larger \mdotz. Particles in less massive disks have larger
stopping times and are therefore less entrained in the gas. Thus, the maximum drift 
velocity moves to smaller particle size with decreasing \mdotz.

\begin{figure} 
\includegraphics[width=6.5in]{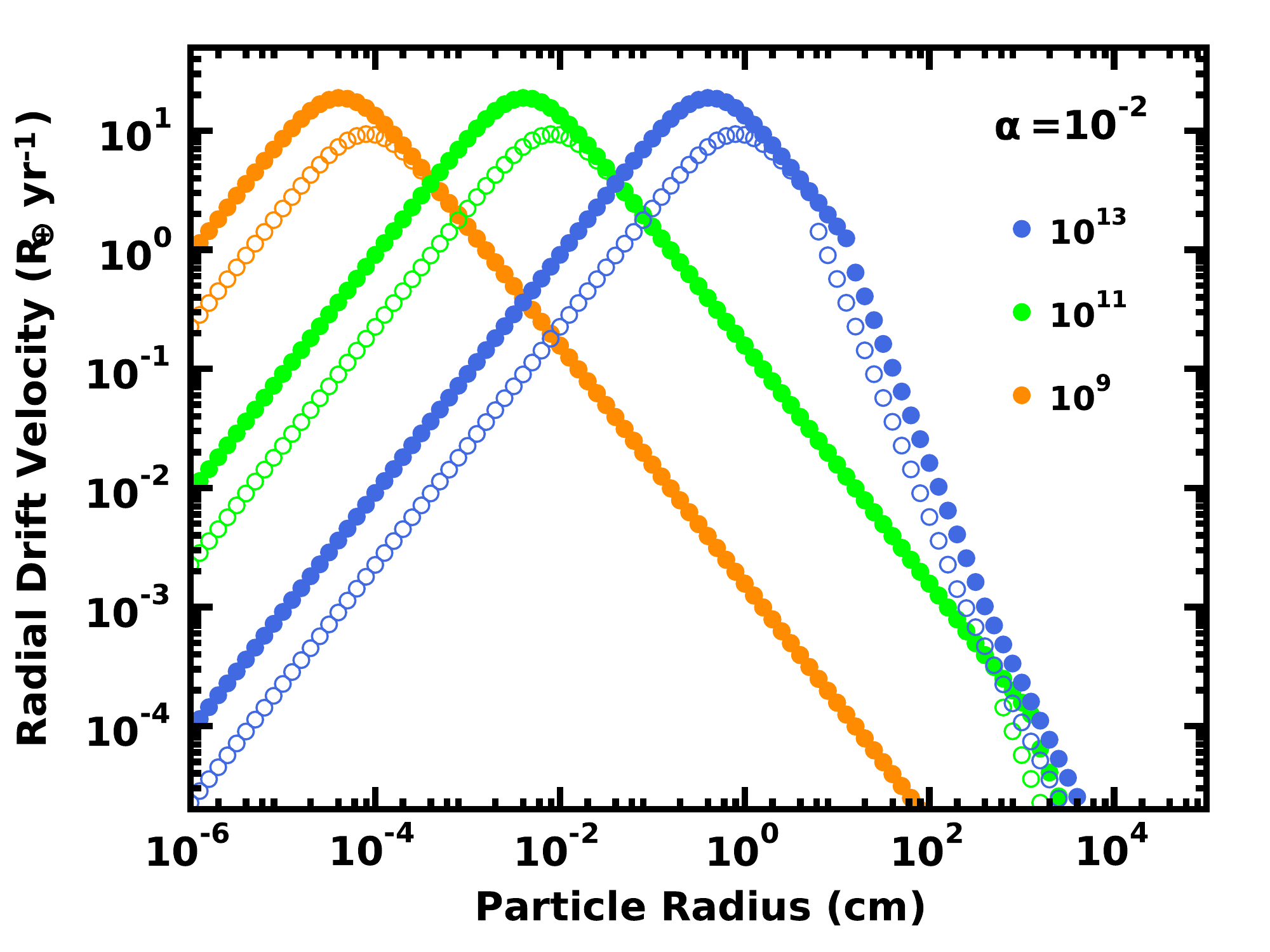}
\vskip 3ex
\caption{
As in Fig.~\ref{fig: drag1} for disks $\alpha = 10^{-2}$. Disks with larger 
$\alpha$ have smaller $\Sigma_g$. Although the maximum drift velocity is 
independent of $\alpha$, smaller particles have the maximum drift velocity
when $\alpha$ is larger.
}
\label{fig: drag2}
\end{figure}

Increasing the disk viscosity parameter $\alpha$ has the same impact as decreasing
\mdotz\ (Fig.~\ref{fig: drag2}). For disks with the same \mdotz\ and \mdotv, those 
with larger $\alpha$ have larger viscosity, lower masses, smaller $\Sigma_g$, and 
larger stopping times \citep[e.g.,][]{lbp1974,weiden1977a}. Compared to 
Fig.~\ref{fig: drag1}, the maximum particle velocity in disks with $\alpha = 10^{-2}$
is shifted to smaller particle sizes. Otherwise, the general variation of the drift
velocity with particle size is fairly independent of $\alpha$.

To derive these results, we assume that the solids and gas are well-mixed, with
$H \lesssim H_g$. In our simulations, however, $H \approx 5 - 10 H_g$. Because the
solids travel through the gas only 10\% to 20\% of each orbit, the actual drift rates 
are 80\% to 90\% smaller than suggested by Figs.~\ref{fig: drag1}--\ref{fig: drag2}.
Thus, the typical maximum drag rate is 2--4~$R_\oplus$~yr$^{-1}$.

Despite these large maximum drift velocities, the drift time is still much longer than
the collision time. With a drift velocity of 2--4~$R_\oplus$~yr$^{-1}$, it takes more
than a decade for a particle to cross our adopted ring of solids. For particles with
$r$ = 0.3--3~cm (\mdotz\ = $10^{13}$~\gs), $r$ = 0.1--1~mm (\mdotz\ = $10^{11}$~\gs), 
or $r$ = 1--10~\mum\ (\mdotz\ = $10^{9}$~\gs), the collision time for a ring with the 
equilibrium mass in solids is $10^3 - 10^4$~s. Thus, the collisional cascade removes
small particles well before gas drag produces a significant radial drift.

In addition to producing a radial drift, the gas damps the orbital $e$ and $\imath$
of particles with $t_s \lesssim T$ \citep{ada1976,weiden1977a}. The damping time scale
is comparable to the time scale for radial drift. With the radial drift time much 
longer than the collision time, the gas cannot reduce the vertical scale height of 
small solids before the collisional cascade grinds them to dust.

Including radiation pressure has a modest impact on this conclusion. Following 
\citet{burns1979}, we define the ratio of the radiation pressure to the local 
gravity:
\begin{equation}
\beta = { { 3 ~ Q_{pr} ~ \lstar } \over { 16 ~ \pi ~ c ~ G ~ r ~ \rho_s ~ \mstar } } ~ ,
\label{eq: beta}
\end{equation}
where $Q_{pr}$ is the Mie scattering coefficient and $c$ is the speed of light. 
For a 0.6~\msun\ white dwarf with \lstar\ $\approx 10^{-2} \lsun$, 
$\beta \approx 0.003 ~ (r / 1\mum)^{-1}$.  With no gas in the system, small grains 
do not respond to radiation pressure before they are vaporized. Once small grains 
are entrained in the gas, however, radiation becomes more important \citep{take2001}. 
With a radial drift velocity of $(\eta - \beta) t_s a \Omega^2$, grains with 
$\beta \gtrsim \eta$ ($\beta \lesssim \eta$) drift radially out (in) through the gas 
\citep{take2001,knb2016}.

For the physical conditions adopted in Figs.~\ref{fig: drag1}--\ref{fig: drag2},
small grains with $r \approx$ 1~\mum\ to 1~mm drift outward at speeds of 
1--100~$R_\oplus$~yr$^{-1}$.  Correcting for the small fraction of time grains 
with large $H$ spend interacting with the gas, it takes grains more than a year 
to move out of our model annulus at 1.15~\rsun.  Compared to the time scale for 
the collisional cascade, these drift rates are still negligible. 

To identify situations where gas drag \textit{is} important, we consider the 
physical conditions in the solids when the time scale for radial drift is 
comparable to the collision time. For simplicity, we set $H \approx H_g$.
From eq.~8 of KB2017, the collision time for a particle with radius $r$ in a 
swarm of solids within our model annulus is
\begin{equation}
t_c \approx 1.6 \times 10^3~{\rm s} 
\left ({ {4 \times 10^{21} ~ {\rm g}} \over M_d } \right ) ~ 
\left ({ r \over {\rm 1~km} } \right ) ~ ,
\label{eq: tc}
\end{equation}
where $M_d$ is the total mass in solids. Setting the collision time to 1~yr yields
the mass where the collision time is roughly equal to the drift time,
$M_d \approx 4 \times 10^{22} (r / {\rm 1~km}) $~g. Setting this mass equal to the 
equilibrium mass of solids in eq.~\ref{eq: meq} allows us to derive the input rate 
of solids, \mdotz, which generates a gas+solid configuration where the collision time 
is comparable to the drift time. This rate is rather large, $\sim 3 \times 10^{17}$~\gs. 
At this rate, the particle size with the maximum drift velocity is roughly 1~km. 
The much smaller accretion rates observed in metallic line white dwarfs precludes 
this option.

We conclude that gas drag has little impact on the evolution of the collisional 
cascade.  For reasonable \mdotz, the collision time is always much shorter than
the radial drift time. The physical conditions that allow the drift time to be
comparable to the collision time are very unlikely.

\subsection{Contacts with Observations}
\label{sec: disc-obs}

In KB2017 and this paper, we have considered collisional cascades as a plausible 
mechanism to place metals in white dwarf photospheres.  However, the lack of robust 
delivery mechanisms and the simplicity of our initial calculations limit our ability 
to explain existing observations. As in KB2017, we list several points of contact 
between current data and theoretical predictions.

\begin{itemize}

\item Circumstellar gas has now been observed in a reasonably large sample of
metallic line white dwarfs \citep[e.g.][]{gansicke2006,gansicke2007,gansicke2008,
melis2010b,farihi2012,melis2012a,debes2012b,wilson2014,manser2016a,hartmann2016,
manser2016b,xu2016,li2017,redfield2017,melis2017}. In some cases, the data place
limits on the surface density of the gas \citep{hartmann2016}. Although other 
models match the observations \citep[e.g.,][]{metzger2012}, cascades with 
\mdotz\ $\approx 10^9 - 10^{11}$ routinely achieve the observed surface densities.

\item Among systems with repeat observations, variable emission (absorption) features 
in the gas are common (rare) \citep[e.g.,][and references therein]{wilson2014,
manser2016a,manser2016b,redfield2017,melis2017}. On one occasion, features appear and 
then slowly fade over time scales of months to years.  More often, consistently visible 
features vary on short time scales. In the cascade picture, occasional large-scale 
collisions can generate a gaseous disk which fades from view before the next collision 
replenishes the disk. During the peak of a cascade, short-term variations in $\Sigma_g$ 
result from stochastic variations in the vaporization rate.

\item In any cascade model, large $\Sigma_g$ and \mdots\ correlate with large $M_d$
and significant IR excesses. Observations appear to support this correlation
\citep[e.g.,][]{manser2016b}. More comprehensive cascade calculations should allow
more robust predictions for comparisons with existing data.

\item Contrary to our assumptions, current observations indicate that gaseous disks 
are not axisymmetric. Large collision events among the solids are not axisymmetric;
proper treatment of the vapor generated in these collisions might produce gaseous
structures similar to those observed.

\item X-ray observations provide an interesting window into the transport of material 
from the Roche limit to the white dwarf photosphere.  From our calculations, we expect 
(i) fairly steady X-ray sources when \mdotv\ and \mdots\ are nearly constant in time 
and (ii) dramatically variable X-ray sources when \r0\ is large and \mdotz\ is small 
(or variable). Unless the incoming gas is channeled by a magnetic field, both sets of 
objects should display the characteristic `flickering' of accreting systems 
\citep[e.g.,][]{sokoloski2003,maoz2015}. Although constraints on the X-ray luminosities 
of metallic line white dwarfs are limited \citep[e.g.,][and references 
therein]{farihi2017b,rappaport2017}, there are some single white dwarfs with 
hard X-rays of unknown origin \citep[e.g.,][]{chu2004,bilikova2010}. Coupled with 
good limits on emission from an IR excess and circumstellar gas, broader surveys for 
X-ray emission among metallic line white dwarfs provide strong tests of theoretical models.

\end{itemize}

\section{SUMMARY}
\label{sec: summary}

We consider the evolution of a gaseous disk fed by a collisional cascade
of solids orbiting
within a narrow ring at the Roche limit of a low-mass white dwarf. In our
picture, solids with radius \r0, eccentricity $e_0 = 0.01$, and inclination 
$\imath_0 = e_0/2$ arrive in the ring at a rate \mdotz. Once the mass in 
solids reaches a critical level, destructive collisions fuel a collisional 
cascade which grinds the solids into 1~\mum\ particles.  Rapid vaporization 
of these small grains produces a ring of metallic gas coincident with the ring
of solids. As the cascade continues, the ring of gas expands into a disk which 
extends from the surface of the white dwarf out past the Roche limit.  The 
disk transports metals originally in the solids onto the white dwarf 
photosphere. 

The evolution of the gaseous disk depends on the properties of the solids (\r0\ and 
\mdotz) and the properties of the gas ($\alpha$, the viscosity parameter, and $T_g$, 
the gas temperature). Systems with \mdotz\ $\gtrsim$ $ \dot{M}_{0, crit}$ $\approx$ 
$3 \times 10^4 ~ (r_0 / {\rm 1~km})^{3.92}$~\gs\ find a steady equilibrium 
state where the vaporization rate \mdotv\ equals \mdotz. Once the ring of gas
spreads into an extended disk, \mdots, the accretion rate from the disk onto 
the central star, is roughly 90\% of \mdotz. With a nearly constant rate of 
mass flow through the disk, the surface density at any distance from the 
central star scales inversely with $\alpha$ and $T_g$. 

When \mdotz\ is smaller, $\lesssim \dot{M}_{0, crit}$, it takes a long time for 
the ring to collect enough solids to begin the collisional cascade. During this 
low state, the vaporization rate \mdotv\ and \mdots\ are much much smaller than 
\mdotz. Once the collisional cascade begins, \mdotv\ and (somewhat later) 
\mdots\ grow very rapidly. During the most intense part of the cascade, 
\mdotv\ and \mdots\ become much larger than \mdotz. The mass in solids then
slowly decreases, reducing \mdotv\ and \mdots. Eventually, the cascade ceases;
\mdotv\ and \mdots\ drop precipitously. As long as the ring continues to accrete
solids at some low rate, the cycle of low and high states repeats.

When a system oscillates between low and high states, $\alpha$ and $T_g$ set the
time scales 
(i) for the newly-formed ring of gas to spread into a disk at the start of the 
cascade, and 
(ii) for the extended disk of gas to drain onto the central star when the cascade 
ends. The time scale for a ring of gas near the Roche limit to spread to the stellar
photosphere is roughly 
$\tau_s \approx 10^3$ $(10^{-3} / \alpha) ~ ({\rm 1500~K} / T_0) $~yr;
the draining time is 5--10 times the spreading time.

As in our previous calculations, the vertical scale height of the solids remains
large throughout the cascade. A revised analysis demonstrates that gas drag has
little impact on the evolution of the solids. Together with results in KB2017, 
it seems unlikely that some combination of collisional, gas dynamical, or radiative
processes can reduce the vertical scale height of the solids, 
$H \approx 10^8 - 10^9$~cm, to levels required in the \citet{jura2003} model,
$H \approx$ 100--1000~cm.

Aside from ultraviolet, optical, and infrared spectroscopy, X-ray fluxes provide 
strong constraints on theoretical models for gaseous disks in metallic line white 
dwarfs. In systems where \r0\ and \mdotz\ vary with time, we expect occasional 
bright states with $L_X / L_\odot \gtrsim 10^{-5}$ and long-lived faint states 
with much smaller $L_X / L_\odot$. By analogy with cataclysmic variables, the 
X-ray temperature during the high state depends on the geometry and optical depth 
of the accreting material \citep[e.g.,][and references therein]{mukai2017}.
Deep surveys with existing \citep[e.g][and references therein]{bilikova2010,
kastner2012} and planned \citep[e.g.,][]{predehl2014} X-ray facilities can test 
these ideas.

\vskip 6ex

We acknowledge generous allotments of computer time on the NASA `discover' cluster.
We thank J. Farihi, M. Geller, M. Payne, S. Rappaport, A. Vanderburg, and D. Veras 
for advice, comments, and encouragement. Comments from an anonymous referee greatly 
improved our presentation.  
Portions of this project were supported by the \textit{NASA } \textit{Outer Planets} 
and \textit{Emerging Worlds} programs through grants NNX11AM37G and NNX17AE24G.

\clearpage

\bibliography{ms.bbl}

\clearpage

\clearpage

\clearpage

\clearpage

\clearpage

\clearpage

\clearpage

\clearpage

\end{document}

%% file: defs.tex
\def\nbody{$n$-body}
\def\deg{\ifmmode {^\circ}\else {$^\circ$}\fi}
\def\degree{\ifmmode {^\circ}\else {$^\circ$}\fi}
\def\mum{\ifmmode {\rm \,\mu {\rm m}}\else $\rm \,\mu {\rm m}$\fi}
\def\arcsec{\ifmmode ^{\prime \prime}\else $^{\prime \prime}$\fi}

\def\inch{\ifmmode ^{\prime \prime}\else $^{\prime \prime}$\fi}
\def\gs{\ifmmode {{\rm g~s^{-1}}}\else ${\rm g~s^{-1}}$\fi}
\def\msunyr{\ifmmode {M_{\odot}~{\rm yr^{-1}}}\else $M_{\odot}~{\rm yr^{-1}}$\fi}
\def\msun{\ifmmode {M_{\odot}}\else $M_{\odot}$\fi}
\def\rsun{\ifmmode {R_{\odot}}\else $R_{\odot}$\fi}
\def\lsun{\ifmmode {L_{\odot}}\else $L_{\odot}$\fi}
\def\mstar{\ifmmode {M_{\star}}\else $M_{\star}$\fi}
\def\rstar{\ifmmode {R_{\star}}\else $R_{\star}$\fi}
\def\tstar{\ifmmode {T_{\star}}\else $T_{\star}$\fi}
\def\lstar{\ifmmode {L_{\star}}\else $L_{\star}$\fi}
\def\mwd{\ifmmode {M_{wd}}\else $M_{wd}$\fi}
\def\rwd{\ifmmode {R_{wd}}\else $R_{wd}$\fi}
\def\twd{\ifmmode {T_{wd}}\else $T_{wd}$\fi}
\def\lwd{\ifmmode {L_{wd}}\else $L_{wd}$\fi}
\def\md{\ifmmode {M_d}\else $M_d$\fi}
\def\ld{\ifmmode {L_d}\else $L_d$\fi}
\def\ad{\ifmmode A_d\else $A_d$\fi}
\def\ldlwd{\ifmmode L_d / L_{wd}\else $L_d / L_{wd}$\fi}
\def\ldlstar{\ifmmode L_d / L_\star\else $L_d / L_{\star}$\fi}
\def\rearth{\ifmmode {\rm R_{\oplus}}\else $\rm R_{\oplus}$\fi}
\def\mearth{\ifmmode {\rm M_{\oplus}}\else $\rm M_{\oplus}$\fi}
\def\qdstar{\ifmmode Q_D^\star\else $Q_D^\star$\fi}
\def\vsqd{\ifmmode v^2 / Q_D^\star\else $v^2 / Q_D^\star$\fi}
\def\kms{\ifmmode {\rm km~s^{-1}}\else $\rm km~s^{-1}$\fi}
\def\ms{\ifmmode {\rm m~s^{-1}}\else $\rm m~s^{-1}$\fi}
\def\vrel{\ifmmode v_{rel}\else $v_{rel}$\fi}
\def\mdot{\ifmmode \dot{M}\else $\dot{M}$\fi}
\def\mdots{\ifmmode \dot{M}_\star\else $\dot{M}_\star$\fi}
\def\mdotv{\ifmmode \dot{M}_v\else $\dot{M}_v$\fi}
\def\mdotz{\ifmmode \dot{M}_0\else $\dot{M}_0$\fi}
\def\mesc{\ifmmode m_{esc}\else $m_{esc}$\fi}
\def\rmin{\ifmmode r_{min}\else $r_{min}$\fi}
\def\rmax{\ifmmode r_{max}\else $r_{max}$\fi}
\def\xmax{\ifmmode x_{max}\else $x_{max}$\fi}
\def\mmin{\ifmmode m_{min}\else $m_{min}$\fi}
\def\mmax{\ifmmode m_{max}\else $m_{max}$\fi}
\def\rmind{\ifmmode r_{min,d}\else $r_{min,d}$\fi}
\def\rmaxd{\ifmmode r_{max,d}\else $r_{max,d}$\fi}
\def\mmaxd{\ifmmode m_{max,d}\else $m_{max,d}$\fi}
\def\vrad{\ifmmode v_{rad}\else $v_{rad}$\fi}
\def\qz{\ifmmode q_{0}\else $q_{0}$\fi}
\def\qi{\ifmmode q_{i}\else $q_{i}$\fi}
\def\ql{\ifmmode q_{l}\else $q_{l}$\fi}
\def\qs{\ifmmode q_{s}\else $q_{s}$\fi}
\def\vhill{\ifmmode v_H\else $r_H$\fi}
\def\rhill{\ifmmode r_H\else $r_H$\fi}
\def\Rhill{\ifmmode R_H\else $R_H$\fi}
\def\rbrk{\ifmmode r_{brk}\else $r_{brk}$\fi}
\def\rdamp{\ifmmode r_{damp}\else $r_{damp}$\fi}
\def\rin{\ifmmode r_{in}\else $r_{in}$\fi}
\def\rout{\ifmmode r_{out}\else $r_{out}$\fi}
\def\tin{\ifmmode t_{in}\else $t_{in}$\fi}
\def\tout{\ifmmode t_{out}\else $t_{out}$\fi}
\def\ain{\ifmmode a_{in}\else $a_{in}$\fi}
\def\aout{\ifmmode a_{out}\else $a_{out}$\fi}
\def\r0{\ifmmode r_{0}\else $r_{0}$\fi}
\def\R0{\ifmmode R_{0}\else $R_{0}$\fi}
\def\m0{\ifmmode m_{0}\else $m_{0}$\fi}
\def\M0{\ifmmode M_{0}\else $M_{0}$\fi}
\def\xm{\ifmmode x_{m}\else $x_{m}$\fi}
\def\sigz{\ifmmode \Sigma_0\else $\Sigma_0$\fi}
\def\ergg{\ifmmode {\rm erg~g^{-1}}\else ${\rm erg~g^{-1}}$\fi}
\def\ergs{\ifmmode {\rm erg~s^{-1}}\else ${\rm erg~s^{-1}}$\fi}
\def\gyr{\ifmmode {\rm g~yr^{-1}}\else ${\rm g~yr^{-1}}$\fi}
\def\cms{\ifmmode {\rm cm~s^{-1}}\else ${\rm cm~s^{-1}}$\fi}
\def\gcms{\ifmmode {\rm g~cm^{-2}}\else $\rm g~cm^{-2}$\fi}
\def\gcmc{\ifmmode {\rm g~cm^{-3}}\else $\rm g~cm^{-3}$\fi}
\def\atil{\ifmmode {\tilde{a}}\else $\tilde{a}$\fi}
\def\ttil{\ifmmode {\tilde{t}}\else $\tilde{t}$\fi}
\def\sqrttt{\ifmmode {\tilde{t}^{1/2}}\else $\tilde{t}^{1/2}$\fi}

\def\orch{{\it Orchestra}}

%% file: ms.bbl
\begin{thebibliography}{}
\expandafter\ifx\csname natexlab\endcsname\relax\def\natexlab#1{#1}\fi

\bibitem[{{Adachi} {et~al.}(1976){Adachi}, {Hayashi}, \& {Nakazawa}}]{ada1976}
{Adachi}, I., {Hayashi}, C., \& {Nakazawa}, K. 1976, Progress of Theoretical
  Physics, 56, 1756

\bibitem[{{Alcock} {et~al.}(1986){Alcock}, {Fristrom}, \&
  {Siegelman}}]{alcock1986}
{Alcock}, C., {Fristrom}, C.~C., \& {Siegelman}, R. 1986, \apj, 302, 462

\bibitem[{{Alcock} \& {Illarionov}(1980)}]{alcock1980b}
{Alcock}, C., \& {Illarionov}, A. 1980, \apj, 235, 541

\bibitem[{{Antoniadou} \& {Veras}(2016)}]{antoniadou2016}
{Antoniadou}, K.~I., \& {Veras}, D. 2016, \mnras, 463, 4108

\bibitem[{{Barber} {et~al.}(2014){Barber}, {Kilic}, {Brown}, \&
  {Gianninas}}]{barber2014}
{Barber}, S.~D., {Kilic}, M., {Brown}, W.~R., \& {Gianninas}, A. 2014, \apj,
  786, 77

\bibitem[{{Barber} {et~al.}(2012){Barber}, {Patterson}, {Kilic}, {Leggett},
  {Dufour}, {Bloom}, \& {Starr}}]{barber2012}
{Barber}, S.~D., {Patterson}, A.~J., {Kilic}, M., {et~al.} 2012, \apj, 760, 26

\bibitem[{{Bath} \& {Pringle}(1981)}]{bath1981}
{Bath}, G.~T., \& {Pringle}, J.~E. 1981, \mnras, 194, 967

\bibitem[{{Bath} \& {Pringle}(1982)}]{bath1982}
---. 1982, \mnras, 199, 267

\bibitem[{{Bear} \& {Soker}(2013)}]{bear2013}
{Bear}, E., \& {Soker}, N. 2013, \na, 19, 56

\bibitem[{{Bergfors} {et~al.}(2014){Bergfors}, {Farihi}, {Dufour}, \&
  {Rocchetto}}]{bergfors2014}
{Bergfors}, C., {Farihi}, J., {Dufour}, P., \& {Rocchetto}, M. 2014, \mnras,
  444, 2147

\bibitem[{{Bil{\'{\i}}kov{\'a}} {et~al.}(2010){Bil{\'{\i}}kov{\'a}}, {Chu},
  {Gruendl}, \& {Maddox}}]{bilikova2010}
{Bil{\'{\i}}kov{\'a}}, J., {Chu}, Y.-H., {Gruendl}, R.~A., \& {Maddox}, L.~A.
  2010, \aj, 140, 1433

\bibitem[{{Bochkarev} \& {Rafikov}(2011)}]{bochkarev2011}
{Bochkarev}, K.~V., \& {Rafikov}, R.~R. 2011, \apj, 741, 36

\bibitem[{{Bonsor} {et~al.}(2017){Bonsor}, {Farihi}, {Wyatt}, \& {van
  Lieshout}}]{bonsor2017}
{Bonsor}, A., {Farihi}, J., {Wyatt}, M.~C., \& {van Lieshout}, R. 2017, \mnras,
  468, 154

\bibitem[{{Bonsor} {et~al.}(2011){Bonsor}, {Mustill}, \& {Wyatt}}]{bonsor2011}
{Bonsor}, A., {Mustill}, A.~J., \& {Wyatt}, M.~C. 2011, \mnras, 414, 930

\bibitem[{{Bonsor} \& {Veras}(2015)}]{bonsor2015}
{Bonsor}, A., \& {Veras}, D. 2015, \mnras, 454, 53

\bibitem[{{Bromley} \& {Kenyon}(2011{\natexlab{a}})}]{bk2011a}
{Bromley}, B.~C., \& {Kenyon}, S.~J. 2011{\natexlab{a}}, \apj, 731, 101

\bibitem[{{Bromley} \& {Kenyon}(2011{\natexlab{b}})}]{bk2011b}
---. 2011{\natexlab{b}}, \apj, 735, 29

\bibitem[{{Bromley} \& {Kenyon}(2013)}]{bk2013}
---. 2013, \apj, 764, 192

\bibitem[{{Bromley} {et~al.}(2012){Bromley}, {Kenyon}, {Geller}, \&
  {Brown}}]{bromley2012}
{Bromley}, B.~C., {Kenyon}, S.~J., {Geller}, M.~J., \& {Brown}, W.~R. 2012,
  \apjl, 749, L42

\bibitem[{{Brown} {et~al.}(2017){Brown}, {Veras}, \& {Gaensicke}}]{brown2017}
{Brown}, J.~C., {Veras}, D., \& {Gaensicke}, B.~T. 2017, \mnras, 468, 1575

\bibitem[{{Burns} {et~al.}(1979){Burns}, {Lamy}, \& {Soter}}]{burns1979}
{Burns}, J.~A., {Lamy}, P.~L., \& {Soter}, S. 1979, Icarus, 40, 1

\bibitem[{{Caiazzo} \& {Heyl}(2017)}]{caiazzo2017}
{Caiazzo}, I., \& {Heyl}, J.~S. 2017, \mnras, 469, 2750

\bibitem[{{Cannizzo} {et~al.}(1990){Cannizzo}, {Lee}, \&
  {Goodman}}]{cannizzo1990}
{Cannizzo}, J.~K., {Lee}, H.~M., \& {Goodman}, J. 1990, \apj, 351, 38

\bibitem[{{Cannizzo} \& {Wheeler}(1984)}]{cannizzo1984}
{Cannizzo}, J.~K., \& {Wheeler}, J.~C. 1984, \apjs, 55, 367

\bibitem[{{Chu} {et~al.}(2004){Chu}, {Guerrero}, {Gruendl}, \&
  {Webbink}}]{chu2004}
{Chu}, Y.-H., {Guerrero}, M.~A., {Gruendl}, R.~A., \& {Webbink}, R.~F. 2004,
  \aj, 127, 477

\bibitem[{{Chu} {et~al.}(2011){Chu}, {Su}, {Bilikova}, {Gruendl}, {De Marco},
  {Guerrero}, {Updike}, {Volk}, \& {Rauch}}]{chu2011}
{Chu}, Y.-H., {Su}, K.~Y.~L., {Bilikova}, J., {et~al.} 2011, \aj, 142, 75

\bibitem[{{Cottrell} \& {Greenstein}(1980)}]{cottrell1980}
{Cottrell}, P.~L., \& {Greenstein}, J.~L. 1980, \apj, 242, 195

\bibitem[{{D'Alessio} {et~al.}(1998){D'Alessio}, {Canto}, {Calvet}, \&
  {Lizano}}]{daless1998}
{D'Alessio}, P., {Canto}, J., {Calvet}, N., \& {Lizano}, S. 1998, \apj, 500,
  411

\bibitem[{{Debes} {et~al.}(2011){Debes}, {Hoard}, {Wachter}, {Leisawitz}, \&
  {Cohen}}]{debes2011}
{Debes}, J.~H., {Hoard}, D.~W., {Wachter}, S., {Leisawitz}, D.~T., \& {Cohen},
  M. 2011, \apjs, 197, 38

\bibitem[{{Debes} {et~al.}(2012{\natexlab{a}}){Debes}, {Kilic}, {Faedi},
  {Shkolnik}, {Lopez-Morales}, {Weinberger}, {Slesnick}, \&
  {West}}]{debes2012b}
{Debes}, J.~H., {Kilic}, M., {Faedi}, F., {et~al.} 2012{\natexlab{a}}, \apj,
  754, 59

\bibitem[{{Debes} \& {Sigurdsson}(2002)}]{debes2002}
{Debes}, J.~H., \& {Sigurdsson}, S. 2002, \apj, 572, 556

\bibitem[{{Debes} {et~al.}(2012{\natexlab{b}}){Debes}, {Walsh}, \&
  {Stark}}]{debes2012a}
{Debes}, J.~H., {Walsh}, K.~J., \& {Stark}, C. 2012{\natexlab{b}}, \apj, 747,
  148

\bibitem[{{Dong} {et~al.}(2010){Dong}, {Wang}, {Lin}, \& {Liu}}]{dong2010}
{Dong}, R., {Wang}, Y., {Lin}, D.~N.~C., \& {Liu}, X.-W. 2010, \apj, 715, 1036

\bibitem[{{Farihi}(2016)}]{farihi2016}
{Farihi}, J. 2016, \nar, 71, 9

\bibitem[{{Farihi} {et~al.}(2012){Farihi}, {G{\"a}nsicke}, {Steele}, {Girven},
  {Burleigh}, {Breedt}, \& {Koester}}]{farihi2012}
{Farihi}, J., {G{\"a}nsicke}, B.~T., {Steele}, P.~R., {et~al.} 2012, \mnras,
  421, 1635

\bibitem[{{Farihi} {et~al.}(2009){Farihi}, {Jura}, \& {Zuckerman}}]{farihi2009}
{Farihi}, J., {Jura}, M., \& {Zuckerman}, B. 2009, \apj, 694, 805

\bibitem[{{Farihi} {et~al.}(2017{\natexlab{a}}){Farihi}, {von Hippel}, \&
  {Pringle}}]{farihi2017a}
{Farihi}, J., {von Hippel}, T., \& {Pringle}, J.~E. 2017{\natexlab{a}}, \mnras,
  471, L145

\bibitem[{{Farihi} {et~al.}(2017{\natexlab{b}}){Farihi}, {Fossati}, {Wheatley},
  {Metzger}, {Mauerhan}, {Bachman}, {G{\"a}nsicke}, {Redfield}, {Cauley},
  {Kochukhov}, {Achilleos}, \& {Stone}}]{farihi2017b}
{Farihi}, J., {Fossati}, L., {Wheatley}, P.~J., {et~al.} 2017{\natexlab{b}},
  ArXiv e-prints, arXiv:1709.08206

\bibitem[{{Frewen} \& {Hansen}(2014)}]{frewen2014}
{Frewen}, S.~F.~N., \& {Hansen}, B.~M.~S. 2014, \mnras, 439, 2442

\bibitem[{{G{\"a}nsicke} {et~al.}(2008){G{\"a}nsicke}, {Koester}, {Marsh},
  {Rebassa-Mansergas}, \& {Southworth}}]{gansicke2008}
{G{\"a}nsicke}, B.~T., {Koester}, D., {Marsh}, T.~R., {Rebassa-Mansergas}, A.,
  \& {Southworth}, J. 2008, \mnras, 391, L103

\bibitem[{{G{\"a}nsicke} {et~al.}(2007){G{\"a}nsicke}, {Marsh}, \&
  {Southworth}}]{gansicke2007}
{G{\"a}nsicke}, B.~T., {Marsh}, T.~R., \& {Southworth}, J. 2007, \mnras, 380,
  L35

\bibitem[{{G{\"a}nsicke} {et~al.}(2006){G{\"a}nsicke}, {Marsh}, {Southworth},
  \& {Rebassa-Mansergas}}]{gansicke2006}
{G{\"a}nsicke}, B.~T., {Marsh}, T.~R., {Southworth}, J., \&
  {Rebassa-Mansergas}, A. 2006, Science, 314, 1908

\bibitem[{{Gezari} {et~al.}(2009){Gezari}, {Heckman}, {Cenko}, {Eracleous},
  {Forster}, {Gon{\c c}alves}, {Martin}, {Morrissey}, {Neff}, {Seibert},
  {Schiminovich}, \& {Wyder}}]{gezari2009}
{Gezari}, S., {Heckman}, T., {Cenko}, S.~B., {et~al.} 2009, \apj, 698, 1367

\bibitem[{{Ghosh} \& {Lamb}(1979)}]{ghosh1979}
{Ghosh}, P., \& {Lamb}, F.~K. 1979, \apj, 232, 259

\bibitem[{{Girven} {et~al.}(2011){Girven}, {G{\"a}nsicke}, {Steeghs}, \&
  {Koester}}]{girven2011}
{Girven}, J., {G{\"a}nsicke}, B.~T., {Steeghs}, D., \& {Koester}, D. 2011,
  \mnras, 417, 1210

\bibitem[{{Hamers} \& {Portegies Zwart}(2016)}]{hamers2016}
{Hamers}, A.~S., \& {Portegies Zwart}, S.~F. 2016, \mnras, 462, L84

\bibitem[{{Hansen} {et~al.}(2006){Hansen}, {Kulkarni}, \&
  {Wiktorowicz}}]{hansen2006}
{Hansen}, B.~M.~S., {Kulkarni}, S., \& {Wiktorowicz}, S. 2006, \aj, 131, 1106

\bibitem[{{Hartmann} {et~al.}(2016){Hartmann}, {Nagel}, {Rauch}, \&
  {Werner}}]{hartmann2016}
{Hartmann}, S., {Nagel}, T., {Rauch}, T., \& {Werner}, K. 2016, \aap, 593, A67

\bibitem[{{Hoard} {et~al.}(2013){Hoard}, {Debes}, {Wachter}, {Leisawitz}, \&
  {Cohen}}]{hoard2013}
{Hoard}, D.~W., {Debes}, J.~H., {Wachter}, S., {Leisawitz}, D.~T., \& {Cohen},
  M. 2013, \apj, 770, 21

\bibitem[{{Holsapple} \& {Michel}(2006)}]{holsapple2006}
{Holsapple}, K.~A., \& {Michel}, P. 2006, \icarus, 183, 331

\bibitem[{{Holsapple} \& {Michel}(2008)}]{holsapple2008}
---. 2008, \icarus, 193, 283

\bibitem[{{Ida} \& {Lin}(2008)}]{ida2008a}
{Ida}, S., \& {Lin}, D.~N.~C. 2008, \apj, 673, 487

\bibitem[{{Jura}(2003)}]{jura2003}
{Jura}, M. 2003, \apjl, 584, L91

\bibitem[{{Jura}(2008)}]{jura2008}
---. 2008, \aj, 135, 1785

\bibitem[{{Jura} {et~al.}(2007{\natexlab{a}}){Jura}, {Farihi}, \&
  {Zuckerman}}]{jura2007b}
{Jura}, M., {Farihi}, J., \& {Zuckerman}, B. 2007{\natexlab{a}}, \apj, 663,
  1285

\bibitem[{{Jura} {et~al.}(2007{\natexlab{b}}){Jura}, {Farihi}, {Zuckerman}, \&
  {Becklin}}]{jura2007a}
{Jura}, M., {Farihi}, J., {Zuckerman}, B., \& {Becklin}, E.~E.
  2007{\natexlab{b}}, \aj, 133, 1927

\bibitem[{{Jura} \& {Young}(2014)}]{jura2014}
{Jura}, M., \& {Young}, E.~D. 2014, Annual Review of Earth and Planetary
  Sciences, 42, 45

\bibitem[{{Kastner} {et~al.}(2012){Kastner}, {Montez}, {Balick}, {Frew},
  {Miszalski}, {Sahai}, {Blackman}, {Chu}, {De Marco}, {Frank}, {Guerrero},
  {Lopez}, {Rapson}, {Zijlstra}, {Behar}, {Bujarrabal}, {Corradi}, {Nordhaus},
  {Parker}, {Sandin}, {Sch{\"o}nberner}, {Soker}, {Sokoloski}, {Steffen},
  {Ueta}, \& {Villaver}}]{kastner2012}
{Kastner}, J.~H., {Montez}, Jr., R., {Balick}, B., {et~al.} 2012, \aj, 144, 58

\bibitem[{{Kennedy} \& {Kenyon}(2008)}]{kenn2008a}
{Kennedy}, G.~M., \& {Kenyon}, S.~J. 2008, \apj, 673, 502

\bibitem[{{Kenyon} \& {Bromley}(2001)}]{kb2001}
{Kenyon}, S.~J., \& {Bromley}, B.~C. 2001, \aj, 121, 538

\bibitem[{{Kenyon} \& {Bromley}(2004)}]{kb2004a}
---. 2004, \aj, 127, 513

\bibitem[{{Kenyon} \& {Bromley}(2008)}]{kb2008}
---. 2008, \apjs, 179, 451

\bibitem[{{Kenyon} \& {Bromley}(2015{\natexlab{a}})}]{kb2015a}
---. 2015{\natexlab{a}}, \apj, 806, 42

\bibitem[{{Kenyon} \& {Bromley}(2015{\natexlab{b}})}]{kb2015b}
---. 2015{\natexlab{b}}, \apj, 811, 60

\bibitem[{{Kenyon} \& {Bromley}(2016)}]{kb2016a}
---. 2016, \apj, 817, 51

\bibitem[{{Kenyon} \& {Bromley}(2017)}]{kb2017b}
---. 2017, \apj, 844, 116

\bibitem[{{Kenyon} {et~al.}(2016){Kenyon}, {Najita}, \& {Bromley}}]{knb2016}
{Kenyon}, S.~J., {Najita}, J.~R., \& {Bromley}, B.~C. 2016, \apj, 831, 8

\bibitem[{{Kenyon} {et~al.}(1988){Kenyon}, {Shipman}, {Sion}, \&
  {Aannestad}}]{kenyon1988}
{Kenyon}, S.~J., {Shipman}, H.~L., {Sion}, E.~M., \& {Aannestad}, P.~A. 1988,
  \apjl, 328, L65

\bibitem[{{Kepler} {et~al.}(2016){Kepler}, {Pelisoli}, {Koester}, {Ourique},
  {Romero}, {Reindl}, {Kleinman}, {Eisenstein}, {Valois}, \&
  {Amaral}}]{kepler2016}
{Kepler}, S.~O., {Pelisoli}, I., {Koester}, D., {et~al.} 2016, \mnras, 455,
  3413

\bibitem[{{Kilic} {et~al.}(2005){Kilic}, {von Hippel}, {Leggett}, \&
  {Winget}}]{kilic2005}
{Kilic}, M., {von Hippel}, T., {Leggett}, S.~K., \& {Winget}, D.~E. 2005,
  \apjl, 632, L115

\bibitem[{{King} {et~al.}(2007){King}, {Pringle}, \& {Livio}}]{king2007}
{King}, A.~R., {Pringle}, J.~E., \& {Livio}, M. 2007, \mnras, 376, 1740

\bibitem[{{Kochanek}(2016)}]{kochanek2016}
{Kochanek}, C.~S. 2016, \mnras, 461, 371

\bibitem[{{Koester} {et~al.}(2014){Koester}, {G{\"a}nsicke}, \&
  {Farihi}}]{koester2014}
{Koester}, D., {G{\"a}nsicke}, B.~T., \& {Farihi}, J. 2014, \aap, 566, A34

\bibitem[{{Koester} \& {Wilken}(2006)}]{koester2006}
{Koester}, D., \& {Wilken}, D. 2006, \aap, 453, 1051

\bibitem[{{Kotko} \& {Lasota}(2012)}]{kotko2012}
{Kotko}, I., \& {Lasota}, J.-P. 2012, \aap, 545, A115

\bibitem[{{Kuulkers} {et~al.}(2006){Kuulkers}, {Norton}, {Schwope}, \&
  {Warner}}]{kuulkers2006}
{Kuulkers}, E., {Norton}, A., {Schwope}, A., \& {Warner}, B. 2006, in Compact
  stellar X-ray sources, ed. W.~H.~G. {Lewin} \& M.~{van der Klis} (Cambridge
  University Press, Cambridge, UK), 421--460

\bibitem[{{Lacombe} {et~al.}(1983){Lacombe}, {Wesemael}, {Fontaine}, \&
  {Liebert}}]{lacombe1983}
{Lacombe}, P., {Wesemael}, F., {Fontaine}, G., \& {Liebert}, J. 1983, \apj,
  272, 660

\bibitem[{{Li} {et~al.}(1998){Li}, {Ferrario}, \& {Wickramasinghe}}]{li1998}
{Li}, J., {Ferrario}, L., \& {Wickramasinghe}, D. 1998, \apjl, 503, L151

\bibitem[{{Li} {et~al.}(2017){Li}, {Zhang}, {Kong}, {Han}, \& {Li}}]{li2017}
{Li}, L., {Zhang}, F., {Kong}, X., {Han}, Q., \& {Li}, J. 2017, \apj, 836, 71

\bibitem[{{Liebert} {et~al.}(1987){Liebert}, {Wehrse}, \&
  {Green}}]{liebert1987}
{Liebert}, J., {Wehrse}, R., \& {Green}, R.~F. 1987, \aap, 175, 173

\bibitem[{{Lodato} \& {Rossi}(2011)}]{lodato2011}
{Lodato}, G., \& {Rossi}, E.~M. 2011, \mnras, 410, 359

\bibitem[{{Lynden-Bell} \& {Pringle}(1974)}]{lbp1974}
{Lynden-Bell}, D., \& {Pringle}, J.~E. 1974, \mnras, 168, 603

\bibitem[{{Lyra} {et~al.}(2010){Lyra}, {Paardekooper}, \& {Mac Low}}]{lyra2010}
{Lyra}, W., {Paardekooper}, S.-J., \& {Mac Low}, M.-M. 2010, \apjl, 715, L68

\bibitem[{{Malamud} \& {Perets}(2016)}]{malamud2016}
{Malamud}, U., \& {Perets}, H.~B. 2016, \apj, 832, 160

\bibitem[{{Malamud} \& {Perets}(2017{\natexlab{a}})}]{malamud2017a}
---. 2017{\natexlab{a}}, \apj, 842, 67

\bibitem[{{Malamud} \& {Perets}(2017{\natexlab{b}})}]{malamud2017b}
---. 2017{\natexlab{b}}, ArXiv e-prints, arXiv:1708.07489

\bibitem[{{Manser} {et~al.}(2016{\natexlab{a}}){Manser}, {G{\"a}nsicke},
  {Koester}, {Marsh}, \& {Southworth}}]{manser2016b}
{Manser}, C.~J., {G{\"a}nsicke}, B.~T., {Koester}, D., {Marsh}, T.~R., \&
  {Southworth}, J. 2016{\natexlab{a}}, \mnras, 462, 1461

\bibitem[{{Manser} {et~al.}(2016{\natexlab{b}}){Manser}, {G{\"a}nsicke},
  {Marsh}, {Veras}, {Koester}, {Breedt}, {Pala}, {Parsons}, \&
  {Southworth}}]{manser2016a}
{Manser}, C.~J., {G{\"a}nsicke}, B.~T., {Marsh}, T.~R., {et~al.}
  2016{\natexlab{b}}, \mnras, 455, 4467

\bibitem[{{Maoz} {et~al.}(2015){Maoz}, {Mazeh}, \& {McQuillan}}]{maoz2015}
{Maoz}, D., {Mazeh}, T., \& {McQuillan}, A. 2015, \mnras, 447, 1749

\bibitem[{{Masset} \& {Papaloizou}(2003)}]{masset2003}
{Masset}, F.~S., \& {Papaloizou}, J.~C.~B. 2003, \apj, 588, 494

\bibitem[{{Melis} \& {Dufour}(2017)}]{melis2017}
{Melis}, C., \& {Dufour}, P. 2017, \apj, 834, 1

\bibitem[{{Melis} {et~al.}(2010){Melis}, {Jura}, {Albert}, {Klein}, \&
  {Zuckerman}}]{melis2010b}
{Melis}, C., {Jura}, M., {Albert}, L., {Klein}, B., \& {Zuckerman}, B. 2010,
  \apj, 722, 1078

\bibitem[{{Melis} {et~al.}(2012){Melis}, {Dufour}, {Farihi}, {Bochanski},
  {Burgasser}, {Parsons}, {G{\"a}nsicke}, {Koester}, \& {Swift}}]{melis2012a}
{Melis}, C., {Dufour}, P., {Farihi}, J., {et~al.} 2012, \apjl, 751, L4

\bibitem[{{Metzger} {et~al.}(2012){Metzger}, {Rafikov}, \&
  {Bochkarev}}]{metzger2012}
{Metzger}, B.~D., {Rafikov}, R.~R., \& {Bochkarev}, K.~V. 2012, \mnras, 423,
  505

\bibitem[{{Meyer} \& {Meyer-Hofmeister}(1982)}]{meyer1982}
{Meyer}, F., \& {Meyer-Hofmeister}, E. 1982, \aap, 106, 34

\bibitem[{{Mineshige} \& {Osaki}(1983)}]{mineshige1983}
{Mineshige}, S., \& {Osaki}, Y. 1983, \pasj, 35, 377

\bibitem[{{Mukai}(2017)}]{mukai2017}
{Mukai}, K. 2017, \pasp, 129, 062001

\bibitem[{{Mustill} {et~al.}(2014){Mustill}, {Veras}, \&
  {Villaver}}]{mustill2014}
{Mustill}, A.~J., {Veras}, D., \& {Villaver}, E. 2014, \mnras, 437, 1404

\bibitem[{{Najita} {et~al.}(2011){Najita}, {{\'A}d{\'a}mkovics}, \&
  {Glassgold}}]{najita2011}
{Najita}, J.~R., {{\'A}d{\'a}mkovics}, M., \& {Glassgold}, A.~E. 2011, \apj,
  743, 147

\bibitem[{{Najita} {et~al.}(2013){Najita}, {Carr}, {Pontoppidan}, {Salyk}, {van
  Dishoeck}, \& {Blake}}]{najita2013}
{Najita}, J.~R., {Carr}, J.~S., {Pontoppidan}, K.~M., {et~al.} 2013, \apj, 766,
  134

\bibitem[{{Nelson} \& {Papaloizou}(2004)}]{nelson2004}
{Nelson}, R.~P., \& {Papaloizou}, J.~C.~B. 2004, \mnras, 350, 849

\bibitem[{{Parriott} \& {Alcock}(1998)}]{parriott1998}
{Parriott}, J., \& {Alcock}, C. 1998, \apj, 501, 357

\bibitem[{{Payne} {et~al.}(2017){Payne}, {Veras}, {G{\"a}nsicke}, \&
  {Holman}}]{payne2017}
{Payne}, M.~J., {Veras}, D., {G{\"a}nsicke}, B.~T., \& {Holman}, M.~J. 2017,
  \mnras, 464, 2557

\bibitem[{{Payne} {et~al.}(2016){Payne}, {Veras}, {Holman}, \&
  {G{\"a}nsicke}}]{payne2016}
{Payne}, M.~J., {Veras}, D., {Holman}, M.~J., \& {G{\"a}nsicke}, B.~T. 2016,
  \mnras, 457, 217

\bibitem[{{Petrovich} \& {Mu{\~n}oz}(2017)}]{petrovich2017}
{Petrovich}, C., \& {Mu{\~n}oz}, D.~J. 2017, \apj, 834, 116

\bibitem[{{Predehl}(2014)}]{predehl2014}
{Predehl}, P. 2014, Astronomische Nachrichten, 335, 517

\bibitem[{{Press} {et~al.}(1992){Press}, {Teukolsky}, {Vetterling}, \&
  {Flannery}}]{press1992}
{Press}, W.~H., {Teukolsky}, S.~A., {Vetterling}, W.~T., \& {Flannery}, B.~P.
  1992, {Numerical recipes in FORTRAN. The art of scientific computing}
  (Cambridge: Cambridge University Press)

\bibitem[{{Pretorius} \& {Knigge}(2012)}]{pretorius2012}
{Pretorius}, M.~L., \& {Knigge}, C. 2012, \mnras, 419, 1442

\bibitem[{{Pringle}(1981)}]{pri1981}
{Pringle}, J.~E. 1981, \araa, 19, 137

\bibitem[{{Rafikov}(2011{\natexlab{a}})}]{rafikov2011a}
{Rafikov}, R.~R. 2011{\natexlab{a}}, \apjl, 732, L3

\bibitem[{{Rafikov}(2011{\natexlab{b}})}]{rafikov2011b}
---. 2011{\natexlab{b}}, \mnras, 416, L55

\bibitem[{{Rafikov} \& {Garmilla}(2012)}]{rafikov2012}
{Rafikov}, R.~R., \& {Garmilla}, J.~A. 2012, \apj, 760, 123

\bibitem[{{Rappaport} {et~al.}(2017){Rappaport}, {Gary}, {Vanderburg}, {Xu},
  {Pooley}, \& {Mukai}}]{rappaport2017}
{Rappaport}, S., {Gary}, B.~L., {Vanderburg}, A., {et~al.} 2017, ArXiv
  e-prints, arXiv:1709.08195

\bibitem[{{Reach} {et~al.}(2005){Reach}, {Kuchner}, {von Hippel}, {Burrows},
  {Mullally}, {Kilic}, \& {Winget}}]{reach2005}
{Reach}, W.~T., {Kuchner}, M.~J., {von Hippel}, T., {et~al.} 2005, \apjl, 635,
  L161

\bibitem[{{Redfield} {et~al.}(2017){Redfield}, {Farihi}, {Cauley}, {Parsons},
  {G{\"a}nsicke}, \& {Duvvuri}}]{redfield2017}
{Redfield}, S., {Farihi}, J., {Cauley}, P.~W., {et~al.} 2017, \apj, 839, 42

\bibitem[{{Rees}(1988)}]{rees1988}
{Rees}, M.~J. 1988, \nat, 333, 523

\bibitem[{{Reis} {et~al.}(2013){Reis}, {Wheatley}, {G{\"a}nsicke}, \&
  {Osborne}}]{reis2013}
{Reis}, R.~C., {Wheatley}, P.~J., {G{\"a}nsicke}, B.~T., \& {Osborne}, J.~P.
  2013, \mnras, 430, 1994

\bibitem[{{Rocchetto} {et~al.}(2015){Rocchetto}, {Farihi}, {G{\"a}nsicke}, \&
  {Bergfors}}]{rocchetto2015}
{Rocchetto}, M., {Farihi}, J., {G{\"a}nsicke}, B.~T., \& {Bergfors}, C. 2015,
  \mnras, 449, 574

\bibitem[{{Shipman} \& {Greenstein}(1983)}]{shipman1983}
{Shipman}, H.~L., \& {Greenstein}, J.~L. 1983, \apj, 266, 761

\bibitem[{{Shipman} {et~al.}(1977){Shipman}, {Greenstein}, \&
  {Boksenberg}}]{shipman1977}
{Shipman}, H.~L., {Greenstein}, J.~L., \& {Boksenberg}, A. 1977, \aj, 82, 480

\bibitem[{{Sion} {et~al.}(1990){Sion}, {Kenyon}, \& {Aannestad}}]{sion1990}
{Sion}, E.~M., {Kenyon}, S.~J., \& {Aannestad}, P.~A. 1990, \apjs, 72, 707

\bibitem[{{Smak}(1999)}]{smak1999}
{Smak}, J. 1999, \actaa, 49, 391

\bibitem[{{Sokoloski} \& {Kenyon}(2003)}]{sokoloski2003}
{Sokoloski}, J.~L., \& {Kenyon}, S.~J. 2003, \apj, 584, 1027

\bibitem[{{Stephan} {et~al.}(2017){Stephan}, {Naoz}, \&
  {Zuckerman}}]{stephan2017}
{Stephan}, A.~P., {Naoz}, S., \& {Zuckerman}, B. 2017, \apjl, 844, L16

\bibitem[{{Stern} {et~al.}(1990){Stern}, {Shull}, \& {Brandt}}]{stern1990}
{Stern}, S.~A., {Shull}, J.~M., \& {Brandt}, J.~C. 1990, \nat, 345, 305

\bibitem[{{Takeuchi} \& {Artymowicz}(2001)}]{take2001}
{Takeuchi}, T., \& {Artymowicz}, P. 2001, \apj, 557, 990

\bibitem[{{Tanaka} {et~al.}(2002){Tanaka}, {Takeuchi}, \& {Ward}}]{tanaka2002}
{Tanaka}, H., {Takeuchi}, T., \& {Ward}, W.~R. 2002, \apj, 565, 1257

\bibitem[{{Tremblay} \& {Bergeron}(2007)}]{tremblay2007}
{Tremblay}, P.-E., \& {Bergeron}, P. 2007, \apj, 657, 1013

\bibitem[{{Veras}(2016)}]{veras2016}
{Veras}, D. 2016, Royal Society Open Science, 3, 150571

\bibitem[{{Veras} {et~al.}(2017){Veras}, {Carter}, {Leinhardt}, \&
  {G{\"a}nsicke}}]{veras2017}
{Veras}, D., {Carter}, P.~J., {Leinhardt}, Z.~M., \& {G{\"a}nsicke}, B.~T.
  2017, \mnras, 465, 1008

\bibitem[{{Veras} {et~al.}(2015{\natexlab{a}}){Veras}, {Eggl}, \&
  {G{\"a}nsicke}}]{veras2015d}
{Veras}, D., {Eggl}, S., \& {G{\"a}nsicke}, B.~T. 2015{\natexlab{a}}, \mnras,
  452, 1945

\bibitem[{{Veras} {et~al.}(2015{\natexlab{b}}){Veras}, {Eggl}, \&
  {G{\"a}nsicke}}]{veras2015b}
---. 2015{\natexlab{b}}, \mnras, 451, 2814

\bibitem[{{Veras} \& {G{\"a}nsicke}(2015)}]{veras2015a}
{Veras}, D., \& {G{\"a}nsicke}, B.~T. 2015, \mnras, 447, 1049

\bibitem[{{Veras} {et~al.}(2014{\natexlab{a}}){Veras}, {Leinhardt}, {Bonsor},
  \& {G{\"a}nsicke}}]{veras2014a}
{Veras}, D., {Leinhardt}, Z.~M., {Bonsor}, A., \& {G{\"a}nsicke}, B.~T.
  2014{\natexlab{a}}, \mnras, 445, 2244

\bibitem[{{Veras} {et~al.}(2013){Veras}, {Mustill}, {Bonsor}, \&
  {Wyatt}}]{veras2013}
{Veras}, D., {Mustill}, A.~J., {Bonsor}, A., \& {Wyatt}, M.~C. 2013, \mnras,
  431, 1686

\bibitem[{{Veras} {et~al.}(2014{\natexlab{b}}){Veras}, {Shannon}, \&
  {G{\"a}nsicke}}]{veras2014c}
{Veras}, D., {Shannon}, A., \& {G{\"a}nsicke}, B.~T. 2014{\natexlab{b}},
  \mnras, 445, 4175

\bibitem[{{Villaver} \& {Livio}(2007)}]{villaver2007}
{Villaver}, E., \& {Livio}, M. 2007, \apj, 661, 1192

\bibitem[{{von Hippel} {et~al.}(2007){von Hippel}, {Kuchner}, {Kilic},
  {Mullally}, \& {Reach}}]{vonhippel2007}
{von Hippel}, T., {Kuchner}, M.~J., {Kilic}, M., {Mullally}, F., \& {Reach},
  W.~T. 2007, \apj, 662, 544

\bibitem[{{Ward}(1997)}]{ward1997}
{Ward}, W.~R. 1997, \icarus, 126, 261

\bibitem[{{Weidenschilling}(1977)}]{weiden1977a}
{Weidenschilling}, S.~J. 1977, \mnras, 180, 57

\bibitem[{{Wilson} {et~al.}(2014){Wilson}, {G{\"a}nsicke}, {Koester}, {Raddi},
  {Breedt}, {Southworth}, \& {Parsons}}]{wilson2014}
{Wilson}, D.~J., {G{\"a}nsicke}, B.~T., {Koester}, D., {et~al.} 2014, \mnras,
  445, 1878

\bibitem[{{Wyatt} {et~al.}(2014){Wyatt}, {Farihi}, {Pringle}, \&
  {Bonsor}}]{wyatt2014}
{Wyatt}, M.~C., {Farihi}, J., {Pringle}, J.~E., \& {Bonsor}, A. 2014, \mnras,
  439, 3371

\bibitem[{{Xu} {et~al.}(2016){Xu}, {Jura}, {Dufour}, \& {Zuckerman}}]{xu2016}
{Xu}, S., {Jura}, M., {Dufour}, P., \& {Zuckerman}, B. 2016, \apjl, 816, L22

\bibitem[{{Xu} {et~al.}(2017){Xu}, {Zuckerman}, {Dufour}, {Young}, {Klein}, \&
  {Jura}}]{xu2017}
{Xu}, S., {Zuckerman}, B., {Dufour}, P., {et~al.} 2017, \apjl, 836, L7

\bibitem[{{Youdin}(2010)}]{youdin2010}
{Youdin}, A.~N. 2010, in EAS Publications Series, Vol.~41, EAS Publications
  Series, ed. {T.~Montmerle, D.~Ehrenreich, \& A.-M.~Lagrange}, 187--207

\bibitem[{{Youdin} \& {Chiang}(2004)}]{youdin2004a}
{Youdin}, A.~N., \& {Chiang}, E.~I. 2004, \apj, 601, 1109

\bibitem[{{Youdin} \& {Kenyon}(2013)}]{youdin2013}
{Youdin}, A.~N., \& {Kenyon}, S.~J. 2013, {From Disks to Planets}, ed. T.~D.
  {Oswalt}, L.~M. {French}, \& P.~{Kalas} (Dordrecht: Springer Science \&
  Business Media), 1

\bibitem[{{Zuckerman} \& {Reid}(1998)}]{zuckerman1998}
{Zuckerman}, B., \& {Reid}, I.~N. 1998, \apjl, 505, L143

\end{thebibliography}
